\documentclass[fleqn,usenatbib,useAMS]{mnras}

\usepackage{graphicx}	
\usepackage{amsmath}	
\usepackage{amssymb}
\usepackage{comment} 

\usepackage{xspace}

\usepackage[dvipsnames]{xcolor}

\newcommand{\dd}{\mathrm{d}}
\newcommand{\vek}[1]{\mathbf{#1}}

\newcommand{\pow}[1]{\ifmmode{}^{#1}\else ${}^{#1}$\fi}

\newcommand{\cm}{\,\ifmmode{{\mathrm{cm}}}\else cm\fi}

\newcommand{\ergps}{\,{\rm erg}\,{\rm s}\ifmmode{}^{-1}\else${}^{-1}$\fi}
\newcommand{\Mpch}{\,{\rm Mpc}\,\ifmmode h^{-1}\else $h^{-1}$\fi}
\newcommand{\snru}{\,\ifmmode{\mathrm{Myr}^{-1}}\else Myr${}^{-1}$\fi}
\newcommand{\kms}{\,\ifmmode{\mathrm{km}\,\mathrm{s}^{-1}}\else km\,s${}^{-1}$\fi\xspace}

\def\lsim{\;\rlap{\lower 2.5pt
   \hbox{$\sim$}}\raise 1.5pt\hbox{$<$}\;}

\newcommand{\cl}{\mathrm{cl}}
\newcommand{\lshatter}{\ifmmode{\ell_{\mathrm{shatter}}}\else $\ell_{\mathrm{shatter}}$\fi}

\usepackage[T1]{fontenc}
\usepackage{ae,aecompl}

\title{Cooling driven coagulation}

\author[Gronke \& Oh]{
Max Gronke${}^{1}$
 and S. Peng Oh${}^{2}$
\\
${}^{1}$ Max Planck Institut für Astrophysik, Karl-Schwarzschild-Straße 1, D-85748 Garching bei München, Germany\\
${}^{2}$ Department of Physics, University of California, Santa Barbara, CA 93106, USA
}

\date{Accepted 2023 June 14. Received 2023 May 24; in original form 2022 October 28}

\pubyear{2023}

\begin{document}
\label{firstpage}
\pagerange{\pageref{firstpage}--\pageref{lastpage}}
\maketitle

\begin{abstract}
Astrophysical gases such as the interstellar-, circumgalactic- or intracluster-medium are commonly multiphase, which poses the question of the structure of these systems. While there are many known processes leading to fragmentation of cold gas embedded in a (turbulent) hot medium, in this work, we focus on the reverse process: coagulation. This is often seen in wind-tunnel and shearing layer simulations, where cold gas fragments spontaneously coalesce. Using 2D and 3D hydrodynamical simulations, we find that sufficiently large ($\gg c_{\rm s} t_{\rm cool}$), perturbed cold gas clouds develop pulsations which ensure cold gas mass growth over an extended period of time ($\gg r / c_{\rm s}$). This mass growth efficiently accelerates hot gas which in turn can entrain cold droplets, leading to coagulation. The attractive inverse square force between cold gas droplets has interesting parallels with gravity; the `monopole' is surface area rather than mass. We develop a simple analytic model which reproduces our numerical findings. 
 \end{abstract}

\begin{keywords}
  galaxies: evolution -- hydrodynamics -- ISM: clouds -- ISM: structure -- galaxies: haloes -- galaxy: kinematics and dynamics
\end{keywords}

\section{Introduction}
\label{sec:intro}

Gases in astrophysics are commonly multiphase, that is, phases with vastly different temperatures exist co-spatially.
We know, for instance, that the interstellar medium is kept in a stable at three phase state due to thermal feedback processes \citep{McKee1977}. More quiescent, the intracluster medium (ICM) or cirgumgalactic medium (CGM) are found to have two main phases, a $T \sim 10^4\,$K `cold' and a $T \gtrsim 10^6\,$K hot phase \citep[e.g.,][]{Tumlinson2017}.
Modeling these gases proves to be extremely difficult due to the corresponding different spatial scales, and large simulations struggle with convergence of the cold gas properties (\citealp{Faucher-Giguere2016,VandeVoort2018,Hummels2018}; also see, e.g., \citealp{2020MNRAS.498.2391N} showing the cold gas covering fractions are unconverged). This is worrisome as this phase corresponds to the fuel for future star-formation and is most commonly compared to observations (e.g., via quasar absorption line studies, \citealp{Crighton2015,Chen2017,Haislmaier2021}, or emission measurements, \citealp{Steidel2011,Hennawi2015,Battaia2018}).
Thus, if a (mis)match to observations is found in such large scale simulations, it is unclear whether this is due to numerics / convergence or whether our understanding of the physical processes is incomplete.
One of the key properties to constrain is therefore a characteristic size of this cold phase where -- hopefully -- one would find convergence in at least the total cold gas mass and other relevant observables. 

Several past and current studies suggested `characteristic length scales' of cold gas \citep{Field1965,McCourt2016,Gronke2018}. Most of them focused on fragmentation processes leading to smaller cold gas `droplets' as a result. Here, we want to focus instead on coagulation between cold gas clouds leading to bigger structures.
\citet{Waters2019} have studied this recently, however -- as we will show below -- in a different regime where the coagulation speed is much slower than the one found in this work.

There are several examples where cooling-induced coagulation appears to be important. For instance: 
\begin{itemize} 
\item{{\it Cloud-Crushing.} In wind-tunnel simulations of an isolated cold cloud subject to a wind, the cloud can initially have a `near death' experience as cloud material is dispersed both streamwise and laterally \citep{Armillotta2017,Gronke2018,Gronnow2018,Li2019a,Kanjilal2020,Farber2021}, particularly for clouds close to the survival radius $r_{\rm crit}$ \citep[cf.][]{Gronke2018,Gronke2020}. As the cloud becomes entrained and shear is reduced, however, cold gas fragments rapidly coagulate back to form a cometary structure. Subsequently, cloud fragments which are peeled off the side of the cloud are refocused back onto the downstream tail.}
\item{{\it Cloud shattering.} In simulations of `cloud-shattering', under-pressured clouds lose sonic contact with their surroundings due to rapid radiative cooling, and are crushed by surrounding hot gas \citep{McCourt2016,Gronke2020}. Since cloud compression overshoots, the cloud subsequently re-expands, and flings small droplets into its surroundings. However, for clouds with a final overdensity (after regaining pressure balance with surroundings) $\chi_{\rm f} \le 300$, the outflowing droplets turn around and coagulate to once again form a monolithic cloud.} 
\item{{\it Turbulence.} In simulations of radiatively cooling multi-phase gas in the presence of extrinsic turbulent driving, coagulation of cold gas clumps are frequent, and play a critical role in maintaining a scale-free power-law distribution ${\rm d}n/{\rm d}M \propto M^{-2}$ \citep{Gronke2022}. While this could simply be geometric (i.e., collisions which occur because clumps are entrained in the turbulent velocity field), there are hints of cooling-induced `focussing'. For instance, we see deviations from this power law at low Mach numbers, which will be presented in future work.} 
\end{itemize}
 
In this work, we want to systematically study the effect of cooling induced coagulation.
This short paper is structured as follows: in Sec.~\ref{sec:methods} we describe our (numerical) methods, in Sec.~\ref{sec:results} we present our results, discuss them in Sec.~\ref{sec:discussion} before we conclude in \S~\ref{sec:conclusion}. Videos visualizing our results can be found at \url{https://max.lyman-alpha.com/coagulation}.

\section{Methods}
\label{sec:methods}

For our hydrodynamical simulation, we use \texttt{Athena} 4.0 \citep{Stone2008} and \texttt{Athena++} \citep{Stone2020}.
We use the HLLC Riemann solver, second-order reconstruction with slope limiters in the primitive variables, and the van Leer unsplit integrator \citep{Gardiner2008}. In both codes, we implemented the \citet{Townsend2009} cooling algorithm which allows for fast and accurate computations of the radiative losses. We adopt a solar metallicity cooling curve to which we fitted a power-law -- similarly to the one used in \citet{McCourt2016} (see their figure 2). As these authors, we use a cooling floor of $T_{\rm floor}=4\times 10^4\,$K. This temperature floor is somewhat high, but in previous work we have shown that mass growth is not sensitive to it \citep{Gronke2018}. We do not employ heating, but in reality the balance between heating and cooling sets this temperature floor.

For this work, we use four different setups:
\begin{itemize}
\item \textit{Isolated cloud.} This three-dimensional setup is similar to the one used in \citep{Gronke2020}, i.e., we placed an isolated cloud of size $\sim r_\cl$\footnote{As the cloud is non-spherical, the effective radius is slightly larger. See \citet{Gronke2020} for details.} with temperature $T_{\rm cl}$ and overdensity $\chi\equiv \rho_\cl / \rho_{\rm h}$ in a hot medium. While the setup is initially in pressure equilibrium (and static), the cloud will (rapidly) cool to $T_{\rm floor}$ leaving $r_\cl$, $\chi$, and $T_{\rm cl}/T_{\rm floor}$ the most important parameters. The purpose of this setup is to systematically study the pulsation induced mass growth discussed in \citep{Gronke2019,Gronke2020,Tan2020}. 
The large perturbation induced by loss of pressure balance with surroundings can occur when a large cloud cools rapidly, or if it is over-run by a shock. Our setup provides a gentler version of the violent loss of pressure balance seen during `shattering' \citep{McCourt2016,Gronke2020}.
\item \textit{Cloud-droplet.} Here, in addition to a cloud as described above, we place a droplet of size $r_{\rm d}$ and temperature $T_{\rm d}$ at a distance $d_0$ away from the cloud. In some cases we also give the droplet an initial velocity $v_{\rm d}$ away from the cloud. The purpose of this setup is to study the coagulation process of the cloud and the droplet. As we need to resolve the droplet sufficiently, we here resort to 2D simulations -- but also carry out 3D ones to study the dimensionality dependence of our results.
\item \textit{Multiple droplets.} We place $N_{\rm d,0}$ droplets with properties as described above randomly within a radius $d$. Again, we perform 2D and 3D simulations with the purpose of studying the coagulation behavior.
\item \textit{Turbulent droplets.} The placement is identical to the 3D `multiple droplets' setup described above but we continuously stir the box in the same manner as in \citet{Gronke2022}, that is, with decaying turbulence as well as continuous driving (to produce a roughly constant kinetic energy) at the scale of the simulation domain with ratio of solenoidal to compressive  components of $\sim 1/3$.
\end{itemize}
For all our setups, we strive to resolve the cold gas by at least $\sim 16$ cells to ensure convergent behavior. We do, however, increase the resolution to $\sim 64$ cells to check this explicitly in some cases (see Appendix~\ref{sec:convergence}; also see \citealp{Tan2020} for an extensive discussion on resolution requirements). Furthermore, we employ `outflowing' boundary conditions -- except in the `turbulent droplets' setup where we used periodic ones.

For the setups involving a droplet, we also inject a advectable scalar field in the droplet furthest away from the origin, which we can use to track droplet motion.

\section{Results}
\label{sec:results}

\begin{figure}
  \centering
  \includegraphics[width=\linewidth]{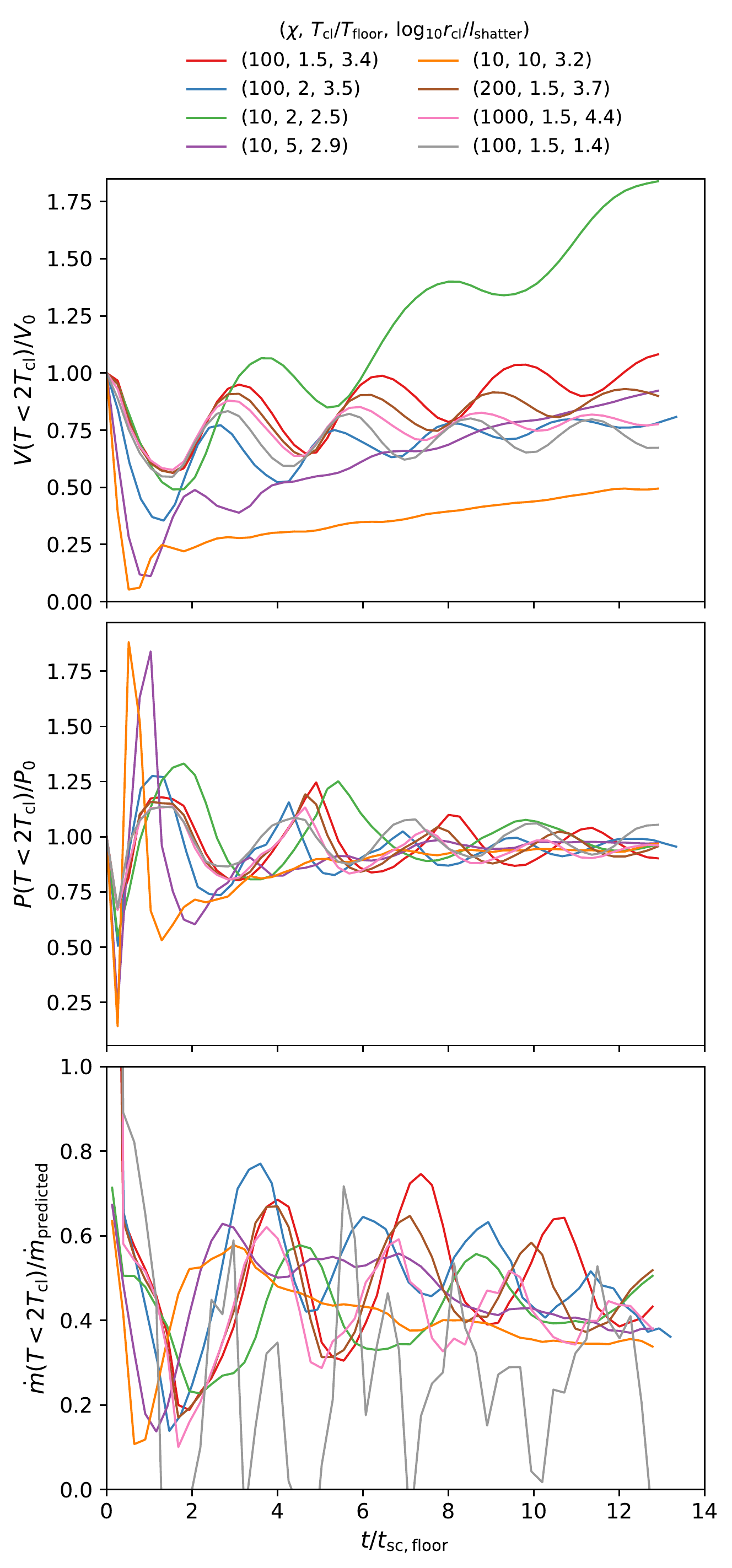}
  \caption{Evolution of pulsating clouds with various initial conditions. The upper panel shows the cold gas volume normalized by its initial value, the central panel the cloud pressure (normalized by the initial / ambient pressure), and the lower panel shows the mass growth rate normalized by the theoretically expected value.}
  \label{fig:mdot_evolution_multiplot}
\end{figure}

\begin{figure}
  \centering
  \includegraphics[width=\linewidth]{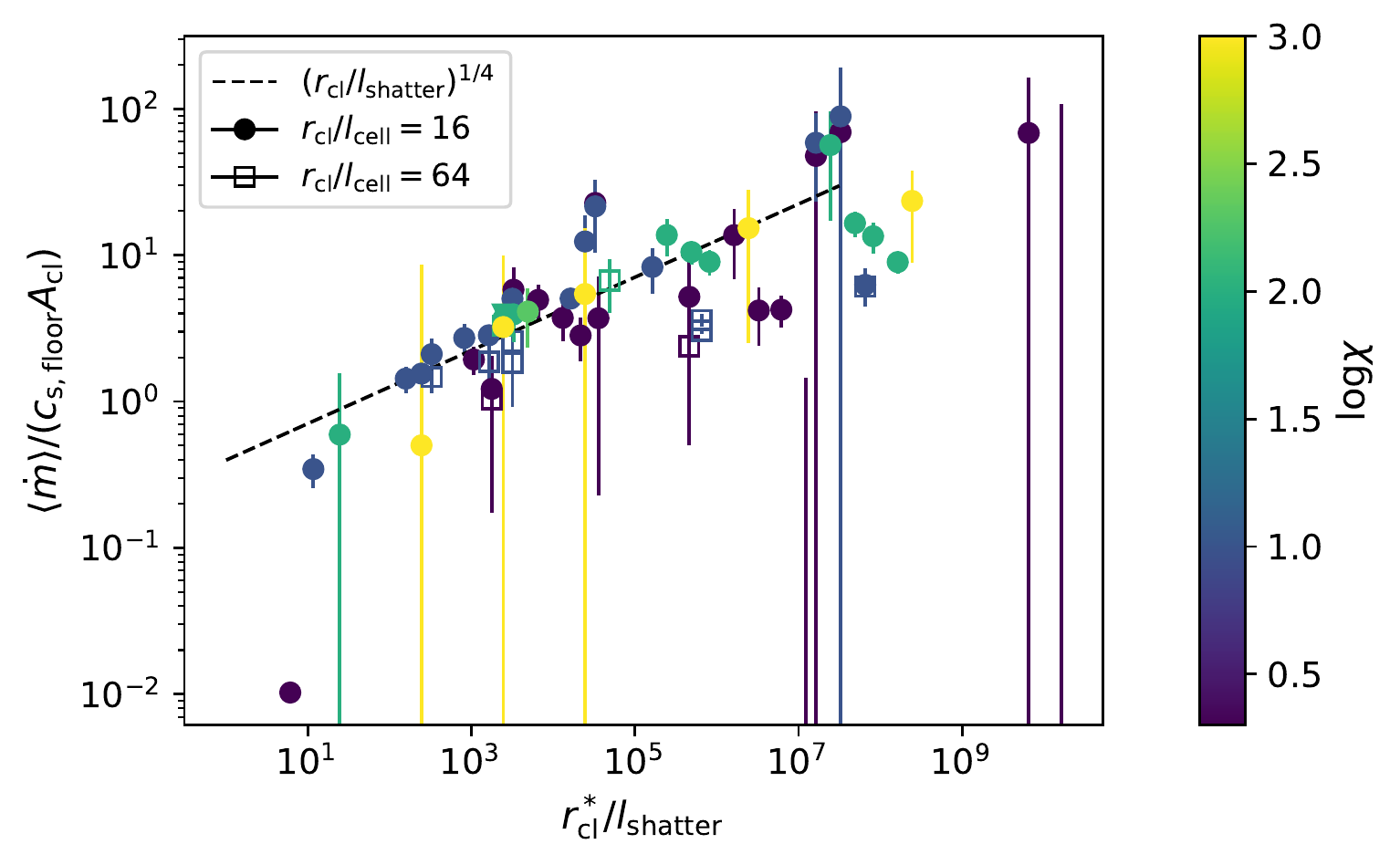}
  \caption{Mass growth rate of different pulsating clouds. Note that we display simulations that do not fragment, i.e., have either $\chi_{\rm final} (10^{-4} r_\cl / \lshatter)^{1/6}\le 300$ or a perturbation of $T_{\rm cl}/T_{\rm floor} < 1.6$. We display simulations with different overdensities $\chi$ (color coded) and resolutions (marker type) and a minimum perturbation of $T_{\rm cl}/T_{\rm floor}>1.1$. The dashed line shows the theoretical expectation, and the error bars correspond to the fluctuation around the median (16th and 84th percentile).}
  \label{fig:mdot_vs_rcl_overview}
\end{figure}

\begin{figure}
  \centering
  \includegraphics[width=\linewidth]{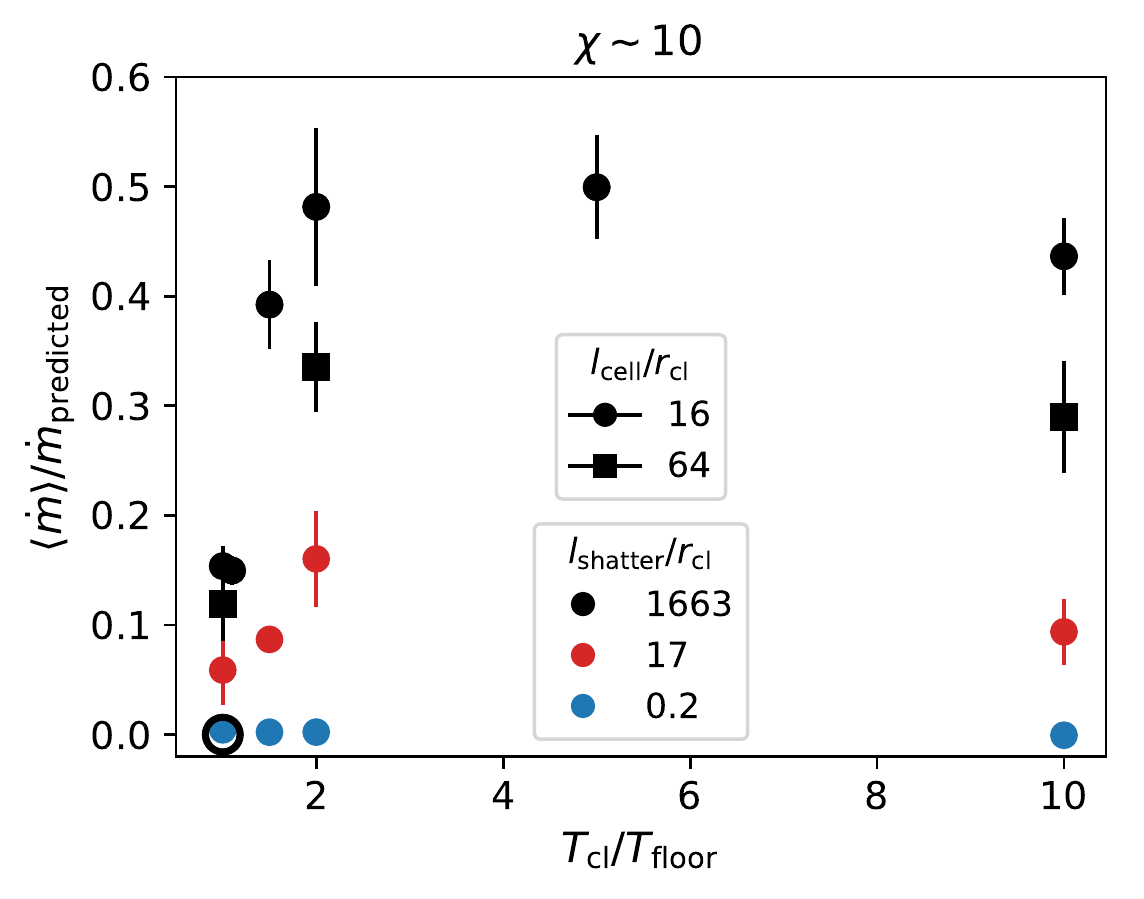}
  \caption{Mass growth versus the initial perturbation $T_{\rm cl} / T_{\rm floor}$. If $r_\cl \gg \lshatter$, $\dot m$ is independent of $T_{\rm cl} / T_{\rm floor}$ for $T_{\rm cl} / T_{\rm floor}\gtrsim 1.5$. The minimum perturbation shown with filled symbols is $T_{\rm cl} / T_{\rm floor}=1.01$. As an unfilled black circle, we also show a simulation with $T_{\rm cl} / T_{\rm floor}=1$ which does not grow.
  }
  \label{fig:mdot_vs_X}
\end{figure}

\subsection{Pulsations \& mass growth in a static medium}
\label{sec:convergence_mdot}

If a cloud does not fragment, it instead oscillates. These oscillations are accompanied by cold gas mass growth\footnote{Note, that the oscillations are crucial in order to obtain a converged mass growth, as we illustrate in Appendix~\ref{sec:osci_conv}.} -- which analogous to our findings in \citet{Gronke2019} we expect to be 
\begin{equation}
  \label{eq:mdot}
  \dot m \sim v_{\mathrm{mix}} A_\cl \rho_{\mathrm{hot}}
\end{equation}
with a cold gas surface area $A_\cl$, and a surrounding hot gas density $\rho_{\mathrm{hot}}$. The characteristic mixing velocity is given by
\begin{equation}
  \label{eq:vmix}
  v_{\mathrm{mix}}\sim \alpha c_{\mathrm{s}} \left( \frac{t_{\mathrm{cool}}}{t_{sc}} \right)^{-1/4}\sim \alpha c_{\mathrm{s}} \left( \frac{r_\cl}{\lshatter} \right)^{1/4},
\end{equation}
where all the quantities $c_{\mathrm{s}}$, $t_{\mathrm{cool}}$, and $t_{\rm sc}$ are evaluated at the floor, that is, $v_{\mathrm{mix}}$ is of the order of the cold gas sound speed, and $\alpha$ is a dimensionless quantity of order unity we calibrate to simulations. This scaling has been confirmed with high-resolution turbulent mixing layer simulations \citep{Tan2020,Fielding2020}\footnote{In general, $v_{\rm mix}\propto u'$ which depends on the geometrical parameters (such as the shear velocity). However, for transonic motion as simulated here, $u'\sim c_{\rm s,cold}$ \citep[see discussion in \S~4.6 and 5.3.3 in ][]{Tan2020}.}.

Figure~\ref{fig:mdot_evolution_multiplot} shows examples of our simulations with different initial overdensities and temperatures ($\chi$, $T_\cl$, respectively), and different cloud sizes. The upper panel shows the cold gas volume from which we see that the oscillations take place on the order of the final sound crossing time $t_{\rm sc, floor}\sim r_\cl / c_{\rm s,floor}$.
The system essentially behaves like a damped, driven oscillator, where damping is due to hydrodynamic drag and driving is due to pressure fluctuations from cooling mixed gas. Initially, there is a transient as the amplitude of the oscillations decay (clearly visible in Fig. \ref{fig:mdot_evolution_multiplot}). However, eventually the system reaches an equilibrium between driving and damping. This is reflected in the fact that mixing induced mass growth is roughly constant for {\it many} sound crossing times (see Fig \ref{fig:mdot_longrun}, where mass growth continues out to $\sim t_{\rm sc,cl}$). Pulsations (and mixing induced mass growth) would cease for a purely damped oscillator. Similar pulsations and long term growth are observed in a cloud accelerated by a wind, even after the cloud is entrained, i.e., the shear between the phases is negligible \citep{Gronke2019,Abruzzo2022}.
In \citet{Gronke2019}, we dubbed these pulsations `overstable sound waves' as they occur on a sound crossing time of the cloud (cf. Fig.~\ref{fig:mdot_evolution_multiplot}).

The lower panel of Fig.~\ref{fig:mdot_evolution_multiplot} shows the mass growth rate (obtained from finite differencing of the cold gas mass) -- which we normalize by the analytic estimate Eq.~\eqref{eq:mdot} (where we used for simplicity the initial cloud size $A_\cl \sim 4 \pi r_\cl^2$). We see that for all the simulations, the values oscillate around $\sim 0.5$, implying $\alpha\sim 0.5$. Moreover, mass growth at this rate keeps this value for many $t_{\rm sc,floor}$, which is longer than we naively expect the initial turbulence in the mixing layer between the hot and cold medium to last. 
Instead, mixing is facilitated and continuously supported by cooling induced pulsations (see also Appendix~\ref{sec:osci_conv} and in particular Fig.~\ref{fig:mdot_longrun} for a longer simulation run).

On overview of the mass growth rate for a range of simulations is shown in Fig.~\ref{fig:mdot_vs_rcl_overview}. Shown are simulations which did not \textit{shatter}, i.e., we excluded the simulations for which the maximum number of droplets was $>100$ which  occurs for $\chi_{\mathrm{final}}= T_\cl / T_{\mathrm{floor}} \chi\gtrsim 300 (10^{-4} r_\cl / \lshatter)^{1/6}$ \citep{Gronke2020}. Note that for this plot we normalized the radii by the theoretical estimate by using the overdensity, temperature, and cloud radius at the point at which the cloud loses sonic contact, i.e., $\chi^* = \chi (r_\cl / r_\cl^*)^3$ with $r_\cl^*=\sqrt{\gamma k_{\rm B} T_\cl^* / (\mu m_{\rm p})} t_{\rm cool}(T^*_\cl, \chi^* \rho_{\mathrm{hot}})$ \citep{Gronke2020}.

Figure~\ref{fig:mdot_vs_rcl_overview} shows that \textit{(i)} the mass growth follows the scaling relation of Eq.~\eqref{eq:vmix} over $\gtrsim 5$ orders of magnitude in cloud size and $\gtrsim 2$ orders of magnitude in overdensity, \textit{(ii)} for small clouds ($r_{\rm cl}^{*} \lesssim 100\lshatter$ for $\chi\gtrsim 100$, larger for smaller overdensities) the mass growth is less than expected, and \textit{(iii)} the high-resolution runs (of $l_{\rm cell} / r_\cl=64$, i.e., a factor of $4$ improvement compared to our fiducial resolution) are consistent with these findings.

As stated above, the clouds in the simulations shown in Fig.~\ref{fig:mdot_vs_rcl_overview} were `sufficiently' perturbed to allow mass growth without shattering (i.e., keeping $\chi_{\rm final}\lesssim 300$ or $T_{\rm cl}/T_{\rm floor}\lesssim 2$). The impact of this initial perturbation -- which sheds light on what `sufficiently' exactly means -- is shown in Fig.~\ref{fig:mdot_vs_X}. In this, we can see that \textit{(i)} as seen before Eq.~\eqref{eq:vmix} is valid only for clouds $r_{\rm cl} \gg \lshatter$ which will pulsate and grow\footnote{Note that Fig.~\ref{fig:mdot_vs_X} shows a small overdensity of only $\chi=10$ which we show to be able to explore a range of $T_\cl / T_{\rm floor}$ values without $\chi_{\rm final}\gtrsim \chi_{\rm crit}$ and, thus, shattering. We also note that $r_\cl / \lshatter$ shown in Fig.~\ref{fig:mdot_vs_X} is clearly a borderline case, thus, falling off the expected $\dot m$.}, \textit{(ii)} if $T_\cl / T_{\mathrm{floor}}\gtrsim 1.5$, the mass growth does not depend on the extent of the perturbation, and \textit{(iii)} for smaller perturbations ($T_{\rm cl}/T_{\rm floor}\lesssim 1.5$), the mass growth does grow with the perturbation but even a value $T_{\rm cl}/T_{\rm floor}\sim 1.01$ (representing our initial random fluctuations, cf.~\S~\ref{sec:methods}) does lead to a significantly larger mass growth than for an unperturbed cloud, where mixing is only due to numerical diffusion.

As mentioned above, the fact that the initial temperature lies above the floor temperature might seem artificial and not occur in nature. However, 
such an abrupt loss of pressure balance can occur in realistic scenarios (e.g., a thermally unstable gas cloud in the ICM/CGM or 
when a cloud is over-run by a shock). When the pressure difference is large, this leads to the well-known `shattering' phenomenon \citep{McCourt2016,Gronke2020}. One can interpret $T_\cl > T_{\mathrm{floor}}$ as a way to simply perturb the system out of pressure balance which happens in reality through such mechanisms. In fact, such oscillations have been observed in simulations where cold gas is ram pressure accelerated; they are seen even in the later, entrained state\citep{Gronke2019,Abruzzo2022}.

\begin{figure}
  \centering
  \includegraphics[width=\linewidth]{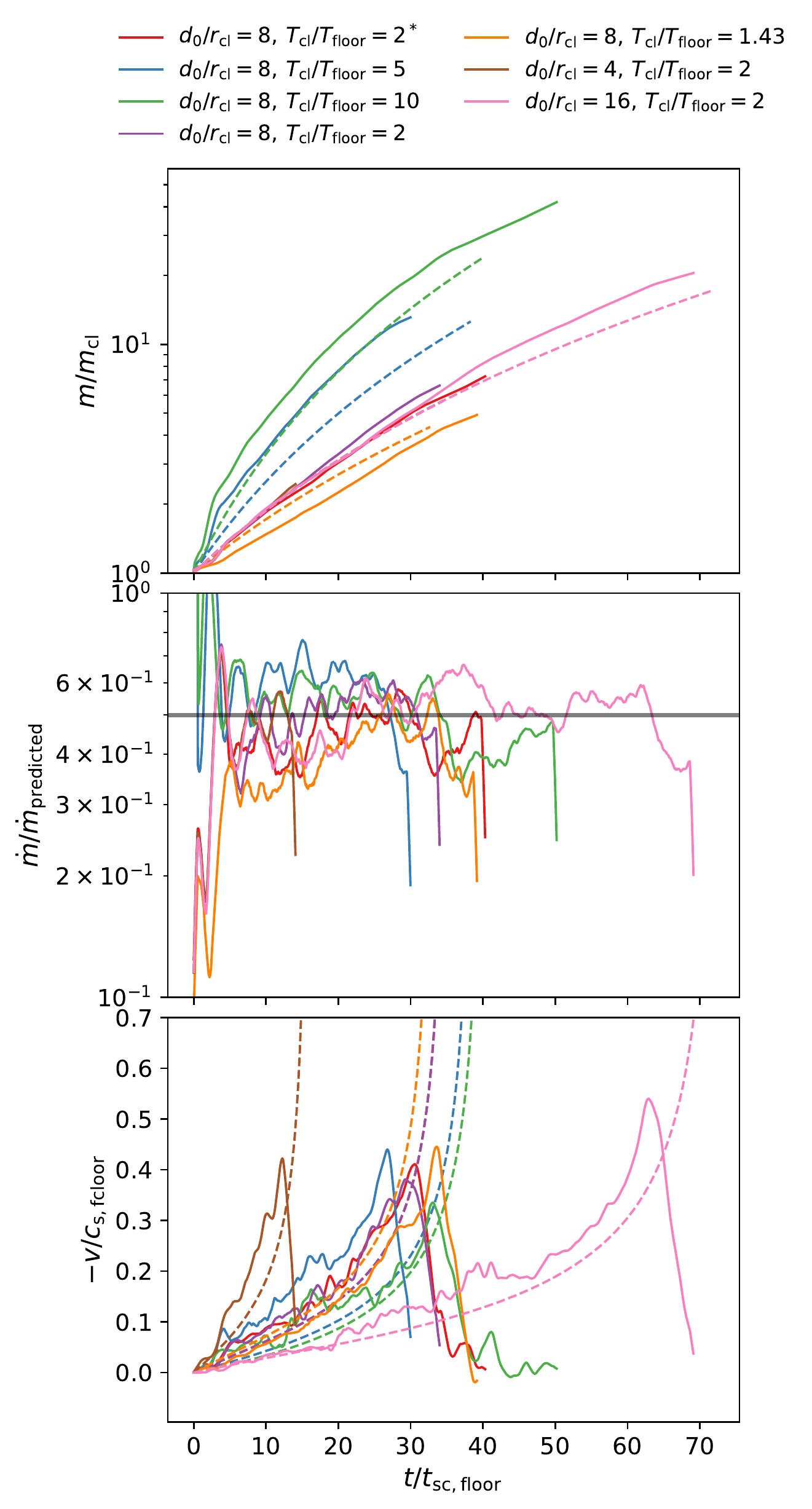}
  \caption{Evolution of $2$D simulations of a droplet located at $d_0/r_\cl$ merging with a cloud of radius $r_\cl$ which cools from $T_\cl$ to $T_{\mathrm{floor}}$. \textit{Top panel:} cold gas mass as a function of time. \textit{Central panel:} Ratio of measured to predicted cold gas mass growth. \textit{Bottom panel:} velocity of the droplet as a function of time. The dashed lines in the upper and lower panel show the curves stemming from solving Eq.~\eqref{eq:eom_mass_growth} with $\alpha=0.5$ which is marked as a black line in the central panel. See \url{https://max.lyman-alpha.com/coagulation} for videos of this setup.}
  \label{fig:coag2d_multiplot}
\end{figure}

\begin{figure}
  \centering
  \includegraphics[width=\linewidth]{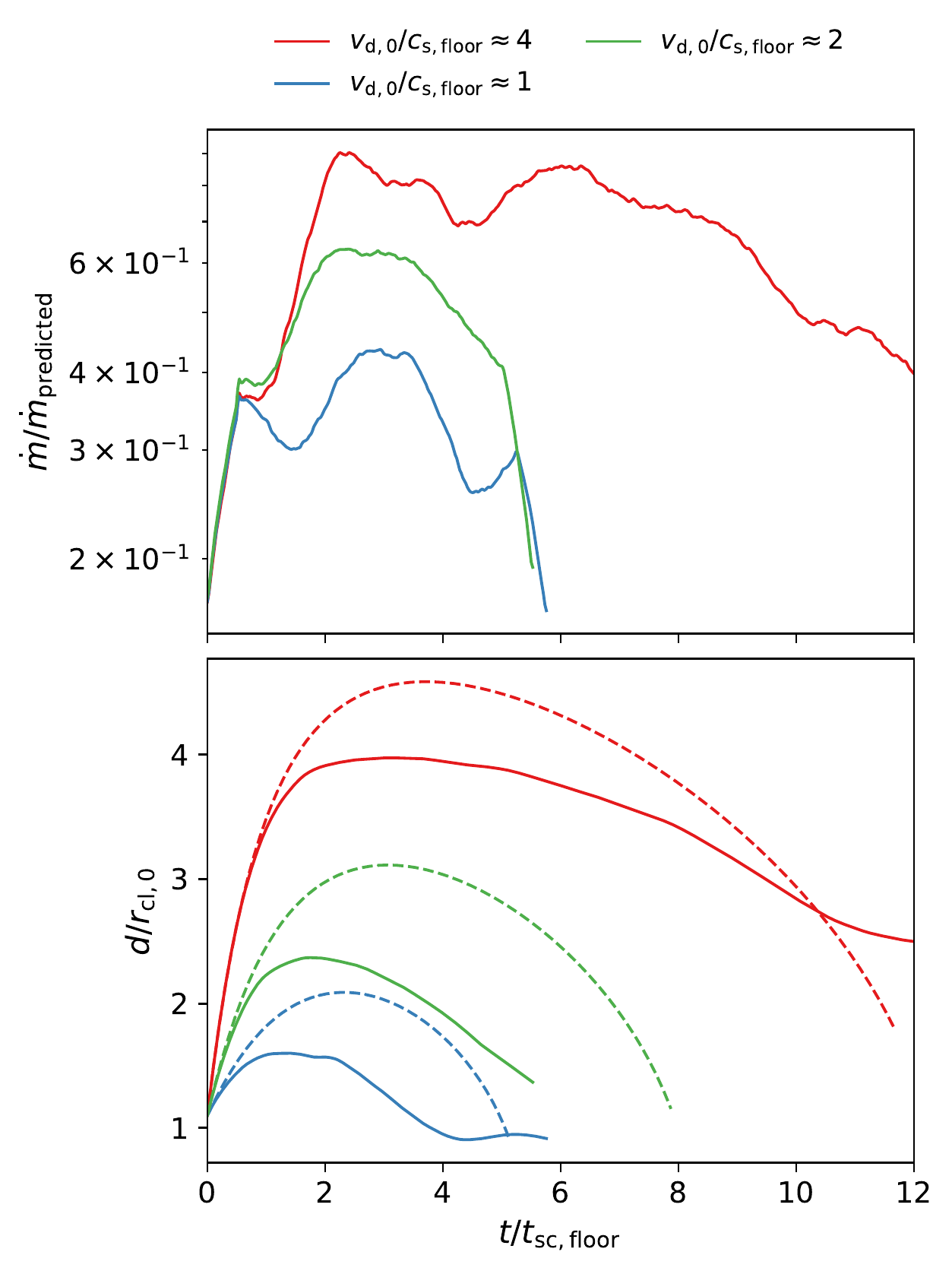}
  \caption{Evolution of $2$D simulations of a droplet located at $d_0/r_\cl=1.1$, and an initial velocity $v_{\rm d,0}$ merging with a cloud of radius $r_\cl$ which cools from $T_\cl\approx 2 T_{\rm floor}$. \textit{Top panel:} mass growth rate normalized by the expected value from Eq.~\eqref{eq:mdot}. \textit{Bottom panel:} location of the droplet as a function of time. The dashed lines correspond to Eq.~\eqref{eq:eom_mass_growth} with a velocity dependent $\alpha_{\rm d}\sim (v_{\rm d} / c_{\rm s,floor})^{1/2}$}
  \label{fig:coag2d_veldrop_multiplot}
\end{figure}

\begin{figure}
  \centering
  \includegraphics[width=\linewidth]{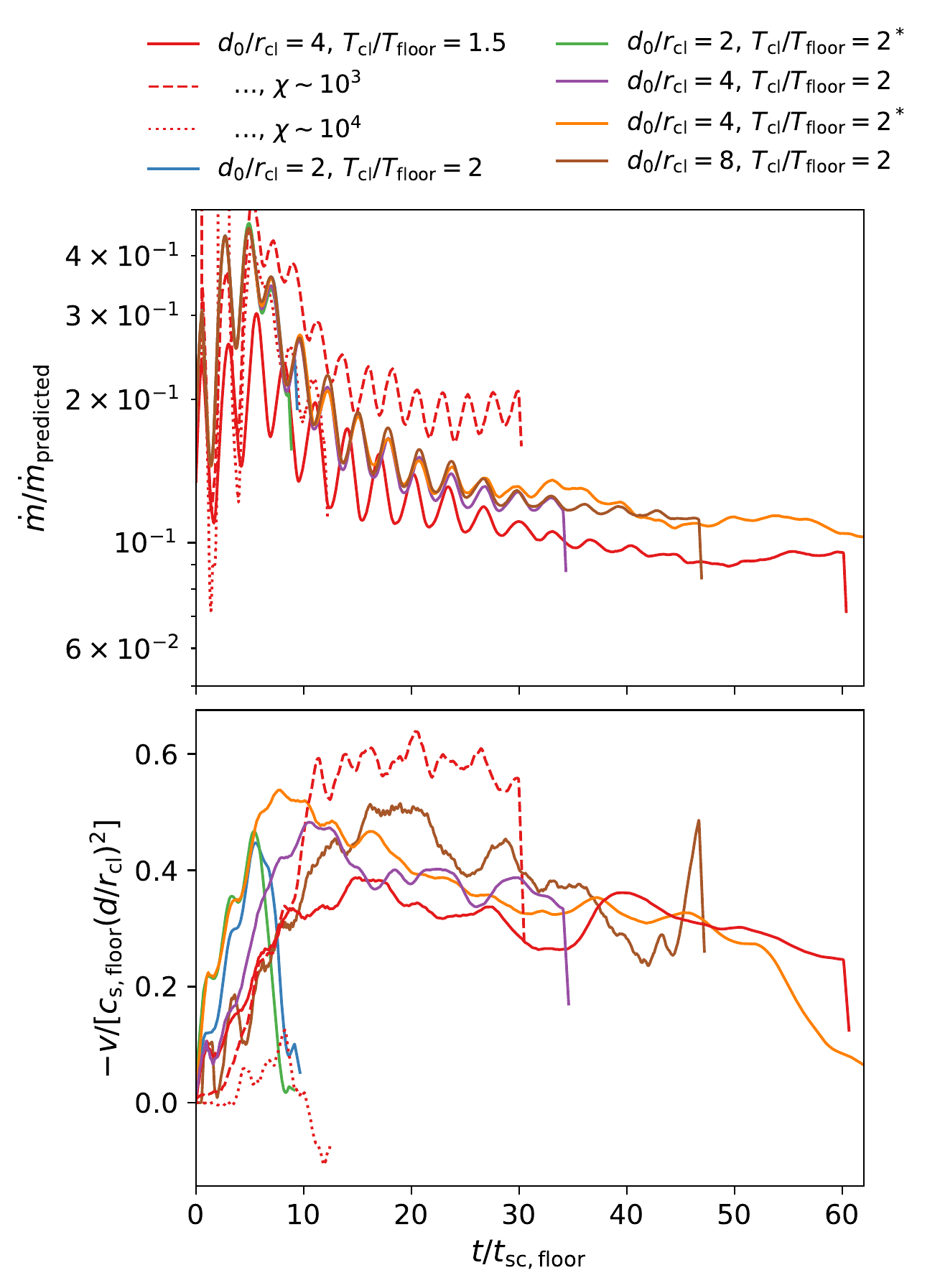}
  \caption{Evolution of $3$D simulations of a droplet located at $d_0/r_\cl$ merging with a cloud of radius $r_\cl$ which cools from $T_\cl$ to $T_{\mathrm{floor}}$. The solid, dashed  and dotted lines show runs with overdensities of $\chi\sim 50$, $\sim 10^3$ and $\sim 10^4$, respectively. The runs marked with ${}^*$ are the ones were we perturbed the droplet, i.e., $T_{\rm d}=T_{\rm cl}$. \textit{Top panel:} ratio of measured to predicted cold gas mass growth. \textit{Bottom panel:} normalized velocity of the droplet as a function of time.}
  \label{fig:coag3d_multiplot}
\end{figure}

\begin{figure*}
  \centering
  \includegraphics[width=\linewidth]{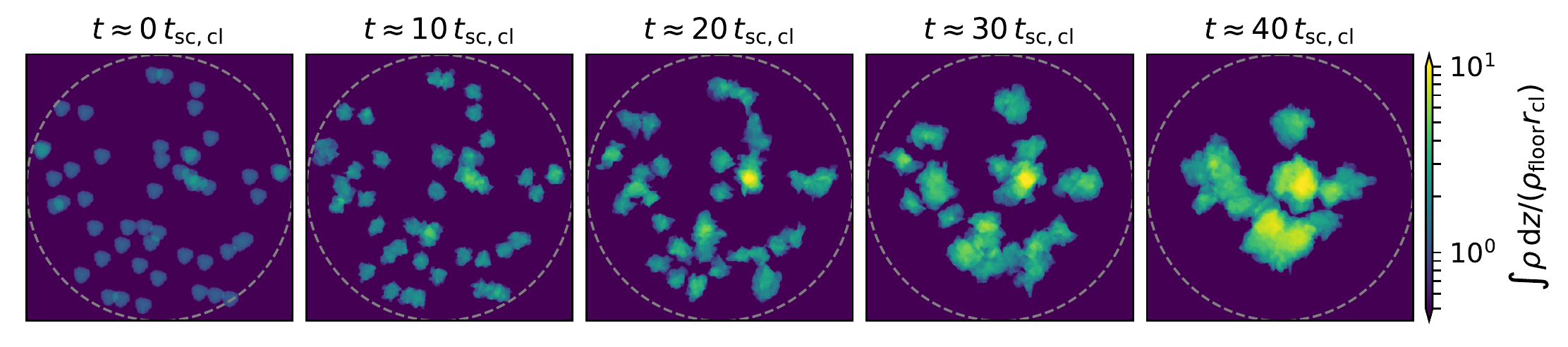}
  \caption{Projections of a $3$D simulations with $N_{\rm d} = 50$ droplets of size $r_{\rm d}\sim 500\lshatter$ placed randomly in a sphere with radius $15 r_{\rm d}$ (marked as white dashed line). The droplets coagulate on a timescale of $\sim 40 t_{\rm sc,cl}$.}
  \label{fig:multi2d_3dcoag}
\end{figure*}

\begin{figure}
  \centering
  \includegraphics[width=\linewidth]{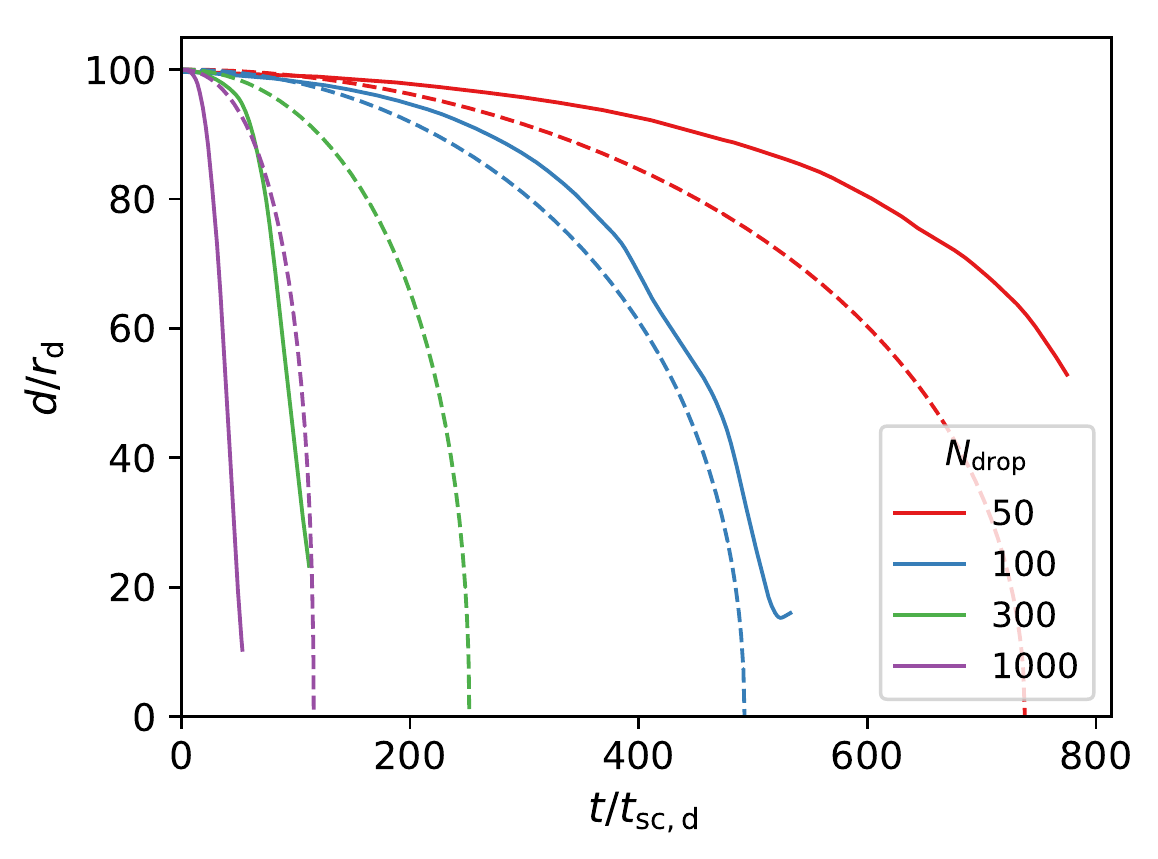}
  \caption{Evolution of $2$D simulations of a fog of droplets randomly placed within $r<100 r_{\rm d}$. Plotted is the distance of the droplet initially furthest away from the origin as a function of time. The dashed lines are our analytical estimate of this scenario using $\alpha=0.1$.}
  \label{fig:coag2d_fog}
\end{figure}

\begin{figure}
  \centering
  \includegraphics[width=\linewidth]{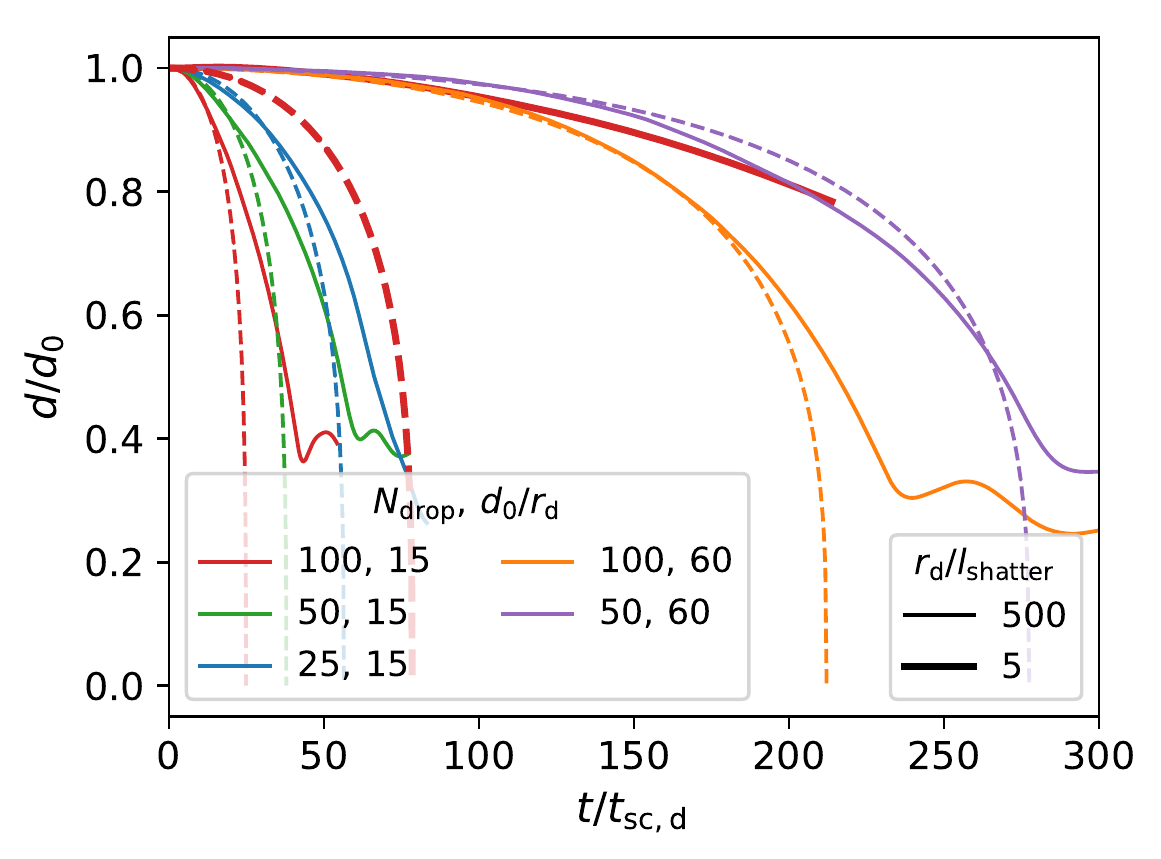}
  \caption{Evolution of $3$D simulations of a fog of droplets randomly places within $r<d_0$. Plotted is the distance of the droplet initially furthest away from the origin as a function of time. The dashed lines are our analytical estimate of this scenario using $\alpha=0.2$. Note how the coagulation is very slow if $r_{\rm d}\lesssim \mathcal{O}(\lshatter)$ due to the lack of pulsations.}
  \label{fig:coag3d_fog}
\end{figure}

\subsection{Cooling induced coagulation}
\label{sec:result_coag}

As we have seen in the previous section, the mass transfer rate from hot to cold medium depends on the size of the cold gas cloud, and is generally an important prediction to compare to observations.  In the circumgalactic medium, for instance, characteristic scales of the cold $\sim 10^4\,$K gas are commonly inferred from absorption line studies \citep[e.g.,][]{Schayelargepopulation2007,Lan2017,Churchill2020} or through emission properties \citep[e.g.,][]{Cantalupo2014,Hennawi2015,Li2021} which indicate the presence of small $\lesssim 100\,$pc clouds. This finding has sparked a range of theoretical studies. As mentioned above, \citet{McCourt2016} suggested droplets of the size of $\lshatter\equiv \mathrm{min}(c_{\rm s}t_{\rm cool})$ to be the outcome of a cooling and fragmentation process. Furthermore, a characteristic size of a cloud $r_{\rm cl}\gtrsim r_{\rm cl,crit} =v_{\rm wind} t_{\rm cool,mix}/\sqrt{\chi}$ is also required for it to survive ram pressure acceleration \citep{Gronke2018,Li2019a,Kanjilal2020}. These predictions can be compared to observations; they also set resolution requirements for larger-scale simulations. Using the example of the circumgalactic medium again, current cosmological simulations are not yet numerically converged in cold gas properties, which makes comparisons to observations problematic \citep[e.g.,][]{Faucher-Giguere2016,Hummels2018}. 

Fragmentation and mixing are processes lowering the size of the cloud. On the other hand, mass growth through cooling (as discussed in the last section), and coagulation of clouds are processes increasing the characteristic cloud size. Coagulation of cold gas clouds is seen to occur in larger scale simulations \citep{Gronke2022}. Here, we want to study the coagulation process due to cooling in highly idealized setups.

\subsubsection{Static, 2D setup}
\label{ssec:2dcoag}
Figure~\ref{fig:coag2d_multiplot} shows the outcome of two-dimensional simulations where we placed a single droplet of size $r_{\rm d}\sim 0.1 r_\cl$ at a distance $d_0$. We perturb the cloud and the droplet as in the previous section by initializing their temperature to $T_\cl > T_{\rm floor}$. As seen before, the cold gas mass growth (upper and central panel of Fig.~\ref{fig:coag2d_multiplot}) follows the expected evolution given by Eq.~\eqref{eq:mdot}.
Due to this mass growth, which is dominated by the cloud, the surrounding hot gas streams towards, and entrains the droplet.
In the lower panel of Fig.~\ref{fig:coag2d_multiplot}, we show the droplet velocity as a function of time.
Note that the droplet gets accelerated both via ram pressure and momentum transfer due to cooling of the mixed material which take place on timescales of $t_{\mathrm{drag}}\sim \chi r_{\mathrm{d}} / v_{\mathrm{hot}}$ and 
\begin{equation}
t_{\mathrm{grow}}\equiv m / \dot m \sim \chi \frac{r}{v_{\rm mix}},
\end{equation} 
respectively.
The ratio of these two timescales is
\begin{equation}
  \frac{t_{\mathrm{drag}}}{t_{\mathrm{grow}}} \sim \frac{v_{\mathrm{mix}}}{v_{\mathrm{hot}}}\sim \frac{d}{r_\cl}
  \label{eq:tdrag_tgrow_ratio}
\end{equation}
where we used $v_{\rm hot}\sim v_{\mathrm{mix}} (r_\cl / d)$ (i.e., assuming the mass growth is dominated by the central cloud; see \S~\ref{ssec:multidrop} for a multidroplet setup), which comes from mass conservation in 2D.  This shows that we expect the momentum transfer via mass growth to dominate.

The net force acting on the droplet, evaluated in the droplet's rest frame, is $F \sim \dot{p} \sim \dot{m} v_{\rm rel} + m \dot{v}_{\rm rel} \sim F_{\rm drag} \sim \rho_{\rm h} v_{\rm rel}^2 A_{\rm cross}$, where the relative velocity between the droplet and the hot wind is $v_{\rm rel}  = v_{\rm mix,cl} (r_{\rm cl}/d) + \dot{d}$ (note that $\dot{d} < 0$). $F_{\rm drag}$ represents the hydrodynamic drag force. We previously saw that $F_{\rm drag}/\dot{m} v \sim t_{\rm grow}/t_{\rm drag} \sim r_{\rm cl}/d \ll 1$ (equation \ref{eq:tdrag_tgrow_ratio}). Thus, the equation of motion simplifies to $m \dot{v}_{\rm rel} \sim - \dot{m} v_{\rm rel}$. Plugging in the expression for $v_{\rm rel}$, this gives: 
\begin{equation}
  \label{eq:eom_mass_growth}
  m(t) \ddot d = -\dot m \left(v_{\mathrm{mix,cl}} \frac{r_\cl}{d} + \dot d \right) + m v_{\rm mix} \left( \frac{r_{\rm cl}}{d^2} \right) \dot{d} 
\end{equation}
with $\dot m\sim 2 \pi v_{\mathrm{mix,d}}r_{\mathrm{d}}\rho_{\mathrm{h}}$ as before. The third term on the right hand side is a fictitious force which arises from the transformation to the non-inertial wind frame (e.g., similar to Coriolis forces). 
Hence, for an entrained droplet, with $t_{\rm grow} \ll t_{\rm adv}$, and $\dot{d}  =  -v_{\rm mix,cl} r_{\rm cl}/d$, the acceleration is given wholly by the third term, $\ddot d = v_{\rm hot}\dot d / d = v_{\rm mix} \left( {r_{\rm cl}}/{d^2} \right) \dot{d}$. The first two terms represent acceleration due to entrainment process, which exerts a force $\sim \dot{m} v_{\rm rel}$.
For completeness, the mass growth of the cloud and the droplets should also be taken into account, 
by integrating $\dot m\sim 2 \pi v_{\mathrm{mix,cl}}r_{\mathrm{cl}}\rho_{\mathrm{h}}$ as well (and analogous for the droplets), and using $r_{\cl}^2\sim m_\cl / (\pi \rho_{\mathrm{cl}})$. 

Note that $v_{\rm mix} \propto r^{1/4}$ is a scale dependent quantity, and thus it is distinct for the cloud and the droplet. However, Eq.~\eqref{eq:vmix} 
was derived in $3$D (and with larger perturbations), so it is unclear if it holds here.
The dashed lines in Fig.~\ref{fig:coag2d_multiplot} shows the outcome of this analytic model and we see that (using $\alpha\sim 0.5$) it fits the numerical solution reasonably well.

\subsubsection{Static, 3D setup including large $\chi$}
Figure~\ref{fig:coag3d_multiplot} shows the evolution of three dimensional runs of the same setup. Note that as here $v_{\rm hot}\propto d^{-2}$ the coagulation process is much slower compared to the 2D runs described above. Nevertheless, the droplets do move towards the cloud and they do so approximately with the velocity expected.

In Fig.~\ref{fig:coag3d_multiplot}, we also show the results of a run with a much larger overdensity of $\chi\sim 1000$ (with dashed lines). We can note that (i) the mass growth follows the predicted scaling Eq.~\eqref{eq:mdot}, (ii) the droplet's motion is independent of $\chi$. This might seem counter-intuitive since the acceleration (both via drag and mass growth) is to first order proportional to $\chi$. However, since 
$t_{\rm grow} \sim \chi r_{\rm d}/v_{\rm mix}$ and $t_{\rm adv} \sim d/[v_{\rm mix}(r_{\rm cl}/d)^2]$, we have: 
\begin{equation} 
\frac{t_{\rm grow}}{t_{\rm adv}} \sim 
3 \left( \frac{\chi}{1000} \right) \left(\frac{r_d/r_{\rm cl}}{0.1}\right)^{3/4} \left(\frac{r_{\rm cl}/d}{0.25}\right)^3
\label{eq:tgrow_tadv}
\end{equation} 
the entrainment time is at most comparable to the advection time (and much shorter for the low $\chi$ case). Thus, the droplet can be treated as comoving with the wind, independent of overdensity.
Interestingly, we find for large $\chi$ ($\gtrsim 10^3$) mass growth rates larger than expected from Eq.~\eqref{eq:tgrow_tadv}. We attribute this to increased fragmentation of the droplet\footnote{We confirm that this fragmentation also occurs in a smooth $v\propto r^{-2}$ background flow, i.e., is not due to perturbations caused by the central cloud.}, which is clearly visible in slice plots. We defer more detailed analysis and better understanding of this boost in mass growth to future work. 

This is no longer the case once the growth time of the droplet is larger than the travel time, i.e., setting $t_{\rm grow,d}\sim t_{\rm travel}$ yields a critical overdensity of 
\begin{equation}
  \label{eq:chi_static}
  \chi_{\rm stuck} \sim \beta \frac{d^3}{r_{\cl}^2 r_{\rm d}} \left( \frac{r_{\rm d}}{r_{\cl}} \right)^{1/4}
\end{equation}
above which the droplet should not move. Here, $\beta$ is a fudge factor encapsulating the deviation from the expected droplet's mass growth rate discussed above.
Setting $\beta\sim 0.1$ (consistent with the mass growths from the simulation), we obtain a $\chi_{\rm stuck}\sim 3600$ (for $d\sim 4 r_{\rm cl}$, $r_\cl / r_{\rm d}\sim 10$). Fig.~\ref{fig:coag3d_multiplot} also shows a simulation with $\chi\sim 10^4$ where indeed the velocity of the droplet $v\sim 0$ (dotted red line in the lower panel of Fig.~\ref{fig:coag3d_multiplot}).

\subsubsection{Static, multidroplet setup}
\label{ssec:multidrop}
Instead of placing a single droplet next to a large cloud, we placed a large number of droplets randomly within a sphere. We again perturb them using an initial temperature of $T/T_{\rm floor}\sim 2$. Due to their combined growth, these droplets will merge to form a single blob. Fig.~\ref{fig:multi2d_3dcoag} visualizes this evolution via density projections of a $3$D simulation. As a proxy of how fast the droplets are coagulating, we use the droplet initially furthest away from the center of the sphere. Fig.~\ref{fig:coag2d_fog} and Fig.~\ref{fig:coag3d_fog} show this droplet's distance to the center of the sphere for two and three-dimensional simulations, respectively.
An increased droplet number density implies more mass growth, and thus faster coagulation. We adopted our cloud-droplet model to this fog of droplets by using a cloud of mass $m_{\rm cl}=N_{\rm drop} m_{\rm drop}$, i.e., considering the combined mass growth.
This simple model (shown as dashed lines in Figs.~\ref{fig:coag2d_fog}, \ref{fig:coag3d_fog}) reproduces the contraction process reasonably well. Discrepancies occur at extremely dense droplet placement when the free-streaming of the hot gas no longer occurs (i.e., shielding becomes important), and for small droplets $r_{\rm d}\lesssim \lshatter$ (thick lines in Fig.~\ref{fig:coag3d_fog}). As shown in \S~\ref{sec:convergence_mdot} for these clouds the pulsations do not occur, leading to slower mass growth -- and hence, the speed of coagulation -- is significantly slower. 

\begin{figure*}
  \centering
  \includegraphics[width=\textwidth]{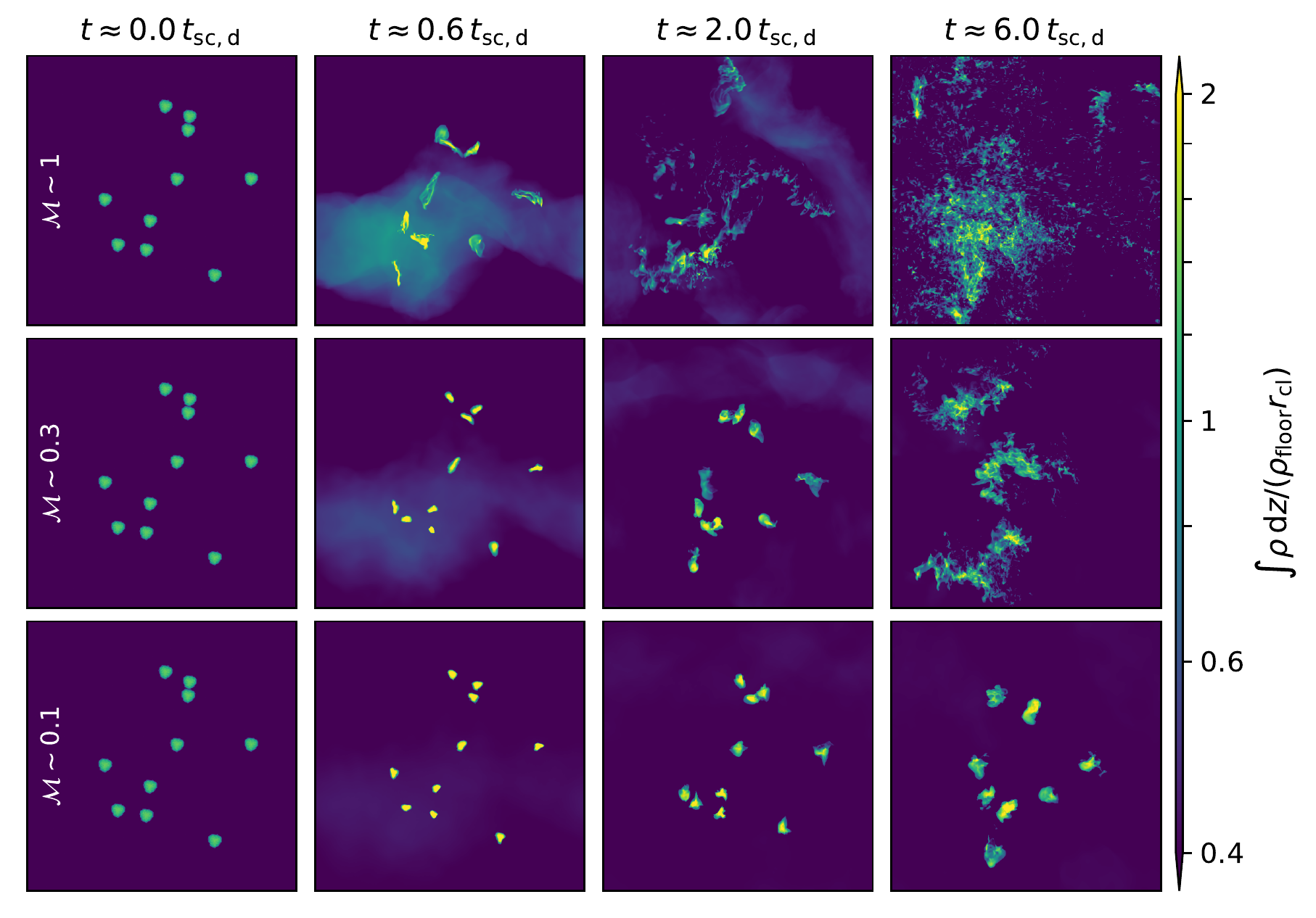}
  \caption{Time evolution of turbulent multiphase boxes with different Mach numbers and $10$ droplets of size $r_{\rm d}\sim 500\lshatter$. While the $\mathcal{M}\sim 1$ case shows fast fragmentation, in the $\mathcal{M}\sim 0.1$ case some droplets have coagulated.}
  \label{fig:3d_turb_coag_density}
\end{figure*}

\subsubsection{Droplets with initial velocity (2D)}
Using this simple model of cooling induced coagulation, we can add additional complexities. Figure~\ref{fig:coag2d_veldrop_multiplot} shows the evolution of 2D runs in which we impose an initial droplet velocity $v_{\rm d,0}$ away from the cloud. This is akin to the situation for `shattering' clouds when droplets fly away with high ($v_{\rm d,0}\lesssim \text{a few}\times c_{\rm s,cold}$) velocities \citep{Gronke2020}. With the model described above, we can reproduce the droplets trajectory quite accurately but note that we use a velocity dependent fudge factor\footnote{In 3D shearing layers, $v_{\rm mix} \sim (u^{\prime})^{3/4} (r/t_{\rm cool})^{1/4}$ scales with the turbulent velocity $u^{\prime}$ rather than the cold gas sound speed, where $u^{\prime} \propto v^{4/5}$, giving $v_{\rm mix} \propto v^{3/5}$ \citep{Tan2020}. Since we have not investigated this in 2D, and also the modification of droplet surface area by the initial velocity, we merely note that this fudge factor (which is $\sim 2$ or less in our numerical experiments) works well.} for the droplet's mass growth rate of $\alpha_{\rm d}\sim (v_{\rm d}/c_{\rm s,floor})^{1/2}$. 
Note that this non-constant $\alpha_{\rm d}$ is inconsistent with the growth used for the (pulsating) cloud thus far (cf. Eq.~\eqref{eq:vmix}). However, the clouds in these simulations undergo significant initial shear and fragmentation, due to its initial velocity. 
As such, it is roughly consistent with findings of higher resolution simulations of turbulent mixing layers showing a dependence of $v_{\rm mix}$ on the shear velocity \citep{Tan2020}.

In summary, the coagulation process of cold gas structures embedded within a hotter surrounding is driven by the cold gas mass growth in two ways. First, in order to sustain the global cold mass growth, the hot medium is moving at a velocity $v \propto v_{\rm mix}(r_{\rm cl}/d)^{2}$ in 3D towards the cold gas. And secondly, due to their own mass growth, droplets become rapidly entrained in this velocity field (cf. Eq.~\eqref{eq:tdrag_tgrow_ratio}). We developed a simple model describing this system, which reproduces our numerical results reasonably well. Such a static setup does not represent, however, reality for most astrophysical systems. We therefore study cold gas mass growth and coagulation in a turbulent setup next.

\begin{figure}
  \centering
  \includegraphics[width=\linewidth]{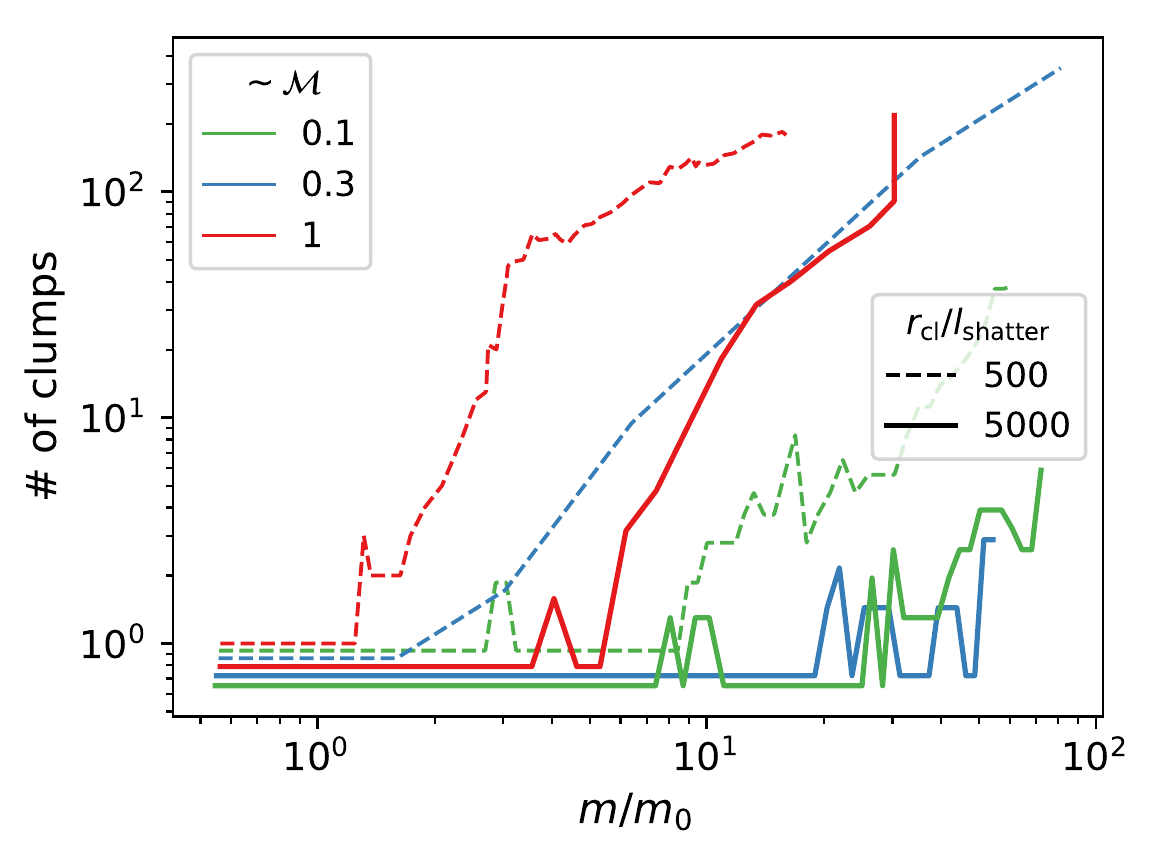}
  \caption{Number of clumps versus the cold gas mass for six simulations of turbulent, multiphase media with different Mach numbers and cold gas sizes. The runs with lower Mach number and larger cloud sizes show stronger coagulation in line with Eq.~\eqref{eq:Mach_coag_numerical}. The lines are horizontally slightly offset for better visualization.}
  \label{fig:turb_nclumps_vs_mass}
\end{figure}

\subsection{Coagulation in a turbulent medium}
\label{sec:res_coag_turb}
As we saw in the last section, the coagulation velocity is $\sim c_{\rm s,c}$, i.e., rather small. In typical astrophysical systems with turbulent velocity dispersion $\sim c_{\rm s,hot}$ it seems at first sight that coagulation cannot `win' over dispersion. This is in line with simulations of multiphase gas in a turbulent medium which show fragmentation of the cold gas \citep[e.g.,][]{Saury2014,Gronke2022,Mohapatra2022a}. However, since the dispersion is not a directed bulk motion like coagulation but instead more akin to a random walk, it is of interest to study the threshold $v_{\rm coag}\sim v_{\rm disp}$. There are two interesting questions: (i) when does a system of clouds coagulate? (ii) when does an individual cloud fragment in the face of turbulence? 

Turbulent dispersion is a large area of research in fluid mechanics \citep[for reviews see, e.g.,][]{Sawford2001,Salazar2009} with a long history. \citet{Batchelor1950} found that initially the mean separation of two particles in a turbulent medium scales as $\langle d^2 \rangle\propto (\epsilon d_0)^{2/3} t^2$ whereas for later times\footnote{Specifically for $t\gg t_{\rm B}\sim d_0^{2/3} \langle \epsilon  \rangle^{-1/3}$ where $d_0$ and $\epsilon$ is the initial separation and the turbulent dissipation, respectively.} the particles `forget' their initial separation and $\langle d^2  \rangle\propto \epsilon t^3$. In both cases, turbulent dispersion is superdiffusive, compared to the customary diffusive expectation $\langle d^2 \rangle \propto t$.

Since we are interested in the dominant process initially -- which governs the further evolution -- we equate the velocity dispersion prior to the ``Batchelor time'' (where the scalings change) to the coagulation velocity.
In this regime, the mean dispersion velocity is given by: 
\begin{equation}
  \label{eq:vdisp}
  \bar v \sim \frac{\dd }{\dd t} \langle d^{2} \rangle^{1/2}x \sim a v_{\rm turb} \left( \frac{d_0}{L} \right)^{1/3}
\end{equation}
where we have used $\epsilon \sim v_{\rm turb}^{3}/L$, $v_{\rm turb}\sim \mathcal{M}c_{\rm c,hot}$ is the driving velocity on the scale of the box and $a\sim 2$ a numerical prefactor\footnote{Specifically, $a=\sqrt{11/6 C_2}$ with $C_2$ being the Kolmogorov constant for the second order velocity structure function. \citet{2013PhRvE..87b3002N} find $C_2\sim 4.02$.}. Note that while Eq.~\eqref{eq:vdisp} follows from the $\langle d^2  \rangle\propto t^2$ scaling described above, it simply represents the Kolmogorov scaling.

If we evaluate turbulence and coagulation at the scale of the cloud $d_0 = r_{\rm cl}$, and require $v_{\rm coag}\sim v_{\rm mix} > \bar{v}$, this yields a critical Mach number:
\begin{align}
  \label{eq:Mach_coag}
  \mathcal{M}_{\rm coag}\sim& \frac{\alpha (r_\cl / \lshatter)^{1/4}}{a (r_\cl / L)^{1/3} \chi^{1/2}}\\
  \sim & 0.16 \left( \frac{r_\cl / \lshatter}{500} \right)^{1/4} \left( \frac{L/r_\cl}{40} \right)^{1/3} \left( \frac{\chi}{100} \right)^{-1/2}
         \label{eq:Mach_coag_numerical}
\end{align}
below which coagulation is stronger than dispersion and clouds should be more robust to fragmentation. 
In Eq.~\eqref{eq:Mach_coag_numerical}, we plugged in typical values and used the fiducial values $a=2$ and $\alpha= 0.2$ as suggested by the result presented in \S~\ref{sec:convergence_mdot} and \citet{Gronke2019}.
In \S\ref{sec:coag_grav}, we also estimate critical Mach numbers below which clouds can coagulate (Eq.~\eqref{eq:Mach_coag_collection2}). The point we will show below is that although there is strong inverse square geometric dimming of coagulation forces, the critical Mach number for coagulation is still $\mathcal{M} \sim v_{\rm mix}/c_{\rm s,h} \sim 0.1 (\chi/100)^{-1/2}$ if cold gas covering fractions are high. 

Figure~\ref{fig:3d_turb_coag_density} shows snapshots of simulations with multiphase, turbulent media. The boxes were initiated with decaying as well as driven turbulence to ensure approximately constant Mach number, and $10$ cold clumps were placed in them (with overdensity $\chi\sim 100$ and size $r_{\rm d}\sim 500\lshatter$; the numerical setup is identical to \citealp{Gronke2022} and we refer the reader to this paper for more details on the setup). The $\mathcal{M}\sim 1$ simulation shows the most fragmentation, whereas in the $\mathcal{M}\sim 0.1$ run, coagulation of droplets occurs.

Figure~\ref{fig:turb_nclumps_vs_mass} shows this behavior in a more quantitative manner. As the turbulent, multiphase medium evolves, the cold gas mass grows (if it is initially larger than some critical size; see \citealp{Gronke2022}) -- and fragments. The extent of this fragmentation depends on the competition between coagulation and dispersion. In Fig.~\ref{fig:turb_nclumps_vs_mass} we show the results from six simulations with different Mach numbers and a single initial cloud of varying size ($256^3$ cells, $L_{\rm box}/r_{\rm cl}=40$, and $\chi\sim 100$) which we analyzed using a clump finding algorithm. In the $r_\cl / \lshatter\sim 500$ case, the $\mathcal{M}\sim 1$ and $\mathcal{M}\sim 0.3$ simulations fragment into $\gtrsim 100$ clumps while in the runs with $r_\cl / \lshatter \sim 5000$ this is only true for $\mathcal{M}\sim 1$. Note that in these simulations, we have kept $L/r_{\rm cl} \sim 40$ constant. Our results are in line with the discussion in \S\ref{sec:coag_grav}. Eq.~\eqref{eq:Mach_coag_numerical} which yields a critical mach number of $\mathcal{M}_{\rm coag}\sim 0.16$ and $0.28$ for the smaller and larger cloud, respectively.

Due to numerical constraints, we can only probe small dynamic temporal and spatial range. However, we showed that coagulation does affect the dynamics of turbulent, multiphase media. Naturally, also other potentially observable properties such as the cloud size distribution are also affected. We will study this point in detail in future work.

\section{Discussion}
\label{sec:discussion}

\subsection{Analogies between coagulation and gravity}
\label{sec:coag_grav}

Consider two clouds separated by a distance $d$. Cloud 1 experiences a force due to cloud 2 given by: 
\begin{equation}
F_{1,2} \sim \dot{m}_1 v_{\rm coag,2} \sim \rho_{\rm h} v_{\rm mix,1} A_1 v_{\rm mix,2} \frac{A_2}{4 \pi d^{2}}
\label{eq:F_1,2} 
\end{equation}
On the other hand, cloud 2 experiences a force due to cloud 1 given by: 
\begin{equation} 
F_{2,1} \sim \dot{m}_2 v_{\rm coag,1} \sim \rho_{\rm h} v_{\rm mix,2} A_2 v_{\rm mix,1} \frac{A_1}{4 \pi d^{2}}\;.
\label{eq:F_2,1} 
\end{equation}
Thus, the two clouds exert equal and opposite attractive forces on one another, with magnitude scaling as the inverse square of their separation $F \propto d^{-2}$. This reminds us of another extremely well-studied force -- gravity -- with the same characteristics, $|F_{1,2}| = |F_{2,1}| \sim G m_1 m_2/d^{2}$. Despite the fact that gravity is relatively `weak'\footnote{For instance, the `gravitational fine-structure constant' $\alpha_{\rm G} \sim G m_p^2/\hbar c \sim 10^{-38}$ is orders of magnitude weaker than the electromagnetic fine-structure constant, $\alpha = e^{2}/\hbar c =1/137$.} and also decays as $F \propto d^{-2}$, it is of course crucial in structuring mass distributions -- despite the simple nature of Newtonian gravity, it gives rise to very rich and complex behavior \citep[e.g.,][]{binney2008}. This is in part because it is a long range attractive force without any shielding --- unlike electromagnetism, there are no negative charges. Similarly, cooling-induced coagulation is a wholly attractive force with no negative charges\footnote{There can be geometric shielding in an optically thick flow (where a cloud blocks hot gas and thus modulates hot gas flow behind it), but we will ignore this complication for now.}. While there are important differences\footnote{For instance, smaller signal speed: coagulational forces propagate at the sound speed of hot gas, and time delay effects can be important.}, the parallels between gravity and coagulation are strong enough to be a useful avenue for thinking about this problem. 

From examining equations \ref{eq:F_1,2} and \ref{eq:F_2,1}, we can identify the analog of gravitational mass to be area $ m \rightarrow A$, and the analog of the gravitational constant to be a peculiar form of kinetic energy density\footnote{Note that $v_{\rm mix} \propto r^{1/4}$ is size dependent. 
We adopt a value $\langle v_{\rm mix} \rangle$ which is understood to be averaged over the size spectrum of cloudlets in the system.}  
$G \rightarrow \rho_{\rm h} v_{\rm mix}^{2}$. Already this tells us about an important difference between gravitational and coagulational dynamics. Mass is conserved under fragmentation and coagulation. However, area is {\it not} conserved: for instance, if one `shatters' a cloud into tiny droplets of radius $r_{\rm d}$, with the number of droplets $N \sim (r_{\rm cl}/r_{\rm d})^{3}$, then the area increases by a factor $N r_{\rm d}^{2} / r_{\rm cl}^{2} \sim r_{\rm cl}/r_{\rm d}$, so that coagulation becomes significantly more important\footnote{As discussed in \S~\ref{sec:convergence_mdot}, this only holds for sizes down to $\sim \lshatter$ after which no pulsations -- and thus no `cooling induced' coagulation -- will occur. However, mixing, cooling (and coagulation) due to external factors such as shear flows can still play a role for these tiny fragments.}. This surface area dependence is key to the strong modulation of coagulation -- `shattering' (which rapidly increases the surface area to mass ratio) boosts the importance of coagulation, while mergers/coagulation itself (which decrease the surface area to mass ratio) reduces the importance of coagulation. In a multi-body system, each cloud is weighted by area, not by mass, and we can follow the motion of an extended distribution by writing an equation for the `center of area' $\langle \mathbf{r}_{\rm CA} \rangle = \int \mathbf{r} \dd A/ \int \dd A$, rather than the center of mass  $\langle \mathbf{r}_{\rm CM} \rangle = \int \mathbf{r} \dd M/ \int \dd M$. We can also think about the analog of the free fall time, $t_{\rm ff} \sim 1/\sqrt{G\rho}$. Consider the total forces acting on a single cloud of mass $m_{\rm cl}$ and area $A_{\rm cl}$ at distance $d$ to the `center of area' of a collection of clouds with total area $A(<d)$ in a sphere with $r=d$, 
\begin{equation} 
m \ddot{d} \sim \rho_{\rm h} v_{\rm mix}^{2} \frac{A_{\rm cl} A_{\rm tot}(<d)}{4 \pi d^{2}} \sim \rho_h v_{\rm mix}^{2} f_A A_{\rm cl}  
\end{equation} 
where $f_A \sim A_{\rm tot}(<d)/4 \pi d^{2}$, 
the number of times a random line of sight with impact parameter less than d to the `center of area intersects a surface\footnote{Similar to optical depth, $f_{\rm A} > 1$ is possible, which boosts the importance of coagulation and decreases $t_{\rm coag}$.}, we can obtain the coagulation time for a cloud embedded in a collection of clouds: 
\begin{equation} 
t_{\rm coag} \sim \left( \frac{\chi}{f_{\rm A}} \right)^{1/2} \frac{(r_{\rm cl} d)^{1/2}}{v_{\rm mix}}.
\label{eq:tcoag}
\end{equation}
Note the appearance of $r_{\rm cl}$ in $t_{\rm coag}$: there will be mass segregation in coagulational collapse, with larger clouds falling to the center more slowly. In gravity, we have the principle of equivalence, due to the equivalence of gravitating and inertial mass: $F=ma = mg$, so $a=g$, independent of mass-- feathers and rocks fall at the same rate in a vacuum. However, for coagulation, $F=ma = m_{\rm coag} g_{\rm coag} \propto A g_{\rm coag}$, so $a \propto A/m \propto 1/r$; larger objects fall more slowly\footnote{Of course, mergers and fragmentation will modulate $t_{\rm coag}$ of a cloud, just as evolving density modulates $t_{\rm ff} \sim 1/\sqrt{G \rho}$.}. 

We can compare the coagulation time Eq.~\eqref{eq:tcoag} with the results shown in Fig.~\ref{fig:coag3d_fog}, where $N_\cl=50$ clouds of size $r_{\rm cl}$ are randomly distributed within a sphere of size $d= 15 r_{\rm cl}$. This gives $f_{\rm A} \approx \int \dd V \, n_{\rm cl} \pi r_{\rm cl}^{2}/(4 \pi r^{2}) \approx N_\cl (r_{\rm cl}/d)^{2}$, where the cloud number density $n_{\cl} \approx 3 N_\cl/(4 \pi d^{3})$. Inserting into Eq.~\eqref{eq:tcoag} (and using for simplicity $v_{\rm mix}\sim c_{\rm s,c}$) yields:
\begin{equation} 
\frac{t_{\rm coag}}{t_{\rm sc,cl}} \sim 80 \left( \frac{\chi}{100} \right)^{1/2} \left( \frac{N_\cl}{50} \right)^{-1/2} \left( \frac{d/r_{\rm cl}}{15} \right)^{3/2}
\end{equation}
which is a factor of 2 larger than the simulation result of ${t_{\rm coag}}/{t_{\rm sc,cl}} \approx 40$. This is good for an order of magnitude estimate, since clouds accelerate as they fall towards the center ($F_{\rm coag} \propto d^{-2}$).  Moreover, Fig. \ref{fig:coag3d_fog} shows rough agreement with a $t_{\rm coag} \propto N_\cl^{-1/2}$ as well as the $t_{\rm coag} \propto d^{3/2}$ scaling. \\

In practice, the clouds, or the hot medium itself, will often be endowed with some relative velocities, which can cause the clouds to disperse. It would be nice to have some rule of thumb or intuition as to when the system coagulates or when it flies apart. In self-gravitating systems, we can compare potential energy $U$ with kinetic energy $K$. If $|U| > K$, the system collapse; if $|U| < K$, it is unbound and flies apart. Could a similar criterion be helpful in coagulating systems? Let us first study how to define potential energy $U$. Consider the work done to separate two clouds from $d_1$ to $d_2$: 
\begin{equation} 
\Delta U = \int_{d_1}^{d_2} F_{\rm coag} \dd r = \frac{\rho_h v_{\rm mix}^{2}}{4 \pi}  A_1 A_2 \left( \frac{1}{d_1} - \frac{1}{d_2} \right).
\end{equation} 
Since $F_{\rm coag}$ is a radial force, it is conservative: $\Delta U$ is independent of the path taken from $d_1$ to $d_2$, and any closed loop (i.e., a path that ends back up at $d_1$ means that no net work\footnote{There {\it will} be work done by other drag forces; we only consider work done by $F_{\rm coag}$.} is done by $F_{\rm coag}$. 

Thus, we can meaningfully define a potential energy $U$ where $F_{\rm coag} = - \nabla U$. If we set $U(\infty) = 0$, we can write: 
\begin{equation} 
U(d) \sim \frac{\rho_{\rm h} v_{\rm mix}^{2} A_1 A_2}{4 \pi d} \sim 3 \rho_h v_{\rm mix}^{2} \Omega_1 \Omega_2 V_d
\end{equation} 
where $V_d \sim (4 \pi/3) d^{3}$, and $\Omega_{i} = A_i/(4 \pi d^{2})$ is the solid angle subtended by cloud $i$. The potential energy density is $\rho v_{\rm mix}^{2} \Omega_1 \Omega_2$: the kinetic energy density $\rho_{\rm h} v_{\rm mix}^{2}$ modulated by the area covering fractions $\Omega_1, \Omega_2$. As the covering fractions increase, so does $|U|$. Thus, fragmentation increases $|U|$, and mergers/coagulation decrease $|U|$.

For a collection of clouds, we sum the potential energy contributions from all pairs of clouds. From the analogy to Newtonian gravity, where $U \sim G M_{\rm tot}^{2}/\langle d \rangle$, where $\langle d \rangle$ is a characteristic scale (such as the half mass radius), we can write the total potential energy as: 
\begin{equation} 
U_{\rm tot} \sim \frac{\rho_{\rm h} v_{\rm mix}^{2}}{4 \pi} \frac{A_{\rm tot}^{2}}{\langle d \rangle}  \sim M_{\rm h} v_{\rm mix}^{2} f_A^2 
\label{eq:U_pot} 
\end{equation}
where $M_{\rm h}\sim \rho_{\rm h} \langle d  \rangle^3$ is the hot gas mass, and the area covering fraction/enhancement factor $f_{\rm A} \sim A_{\rm tot}/\langle d \rangle^{2}$ modulates the strength of potential energy. Thus, if $f_{\rm A} > 1$ (and indeed, $f_{\rm A} \gg 1$ is possible in 'fog-like' cloud topology), the potential energy will exceed the naive bound $M_{\rm h} v_{\rm mix}^{2}$, due to the superposition of the flows from multiple small clouds. Of course, at that point a more careful treatment which takes geometric shielding into account is necessary.

What about the kinetic energy? There are at least two classes of problems: (i) the hot gas is initially static and the cloudlets have some initial relative velocity. A prototypical example is cloud shattering. This statement is also approximately true of the cloud growth problem in the frame of the wind, when cloud fragments of different size have undergone differential acceleration. In this case the relevant kinetic energy is $K \sim M_c \sigma_c^2$, where $\sigma_c^2(d)$ is the velocity dispersion of cold gas at scale $d$. (ii) The hot gas velocity field has significant velocity structure (e.g., in the form of shear or turbulence), and can potentially entrain the clouds. In this case, the relevant kinetic energy is $K \sim M_h \sigma_h^2$, where $\sigma_h^2$ is the velocity dispersion of hot gas. 

Although energy is not strictly conserved\footnote{In the first case, the hot medium provides an additional drag force which slows dispersal and promotes coagulation. In the second case, the hot medium (if it entrains the clouds) promotes dispersal. Therefore, unlike the self-gravitating case, there are additional dissipative or driving forces acting, besides the conservative force. Thus, there is no energy conservation: in the first case, kinetic energy decays (due to `friction' against the hot gas); in the second case, cloud entrainment transfers kinetic energy from the hot to cold gas. And, as previously noted, fragmentation/mergers modulates potential energy.}, we can use this to estimate whether coagulation is likely to happen. For coagulation to happen, we require that $|U_{\rm tot}| > |K_{\rm tot}|$, or $\sigma_h(d) < v_{\rm mix} f_A$. If we use Kolmogorov scalings for $\sigma_h(d)$, this gives a critical Mach number for coagulation:  
\begin{align}
  \label{eq:Mach_coag_collection1}
  \mathcal{M}_{\rm coag}\sim& \frac{\alpha (r_\cl / \lshatter)^{1/4} f_A }{(r_\cl / L)^{1/3} \chi^{1/2}}
\end{align}
below which clouds will coagulate. 
This yields the same scalings as found in Eq.~\eqref{eq:Mach_coag}, but a much lower normalization, due to the small value of $f_{\rm A} \sim 0.05 [(r_{\rm cl}/d)/15]^2 (N_{\rm cl}/10)$. 
We discuss this in \S~\ref{sec:coag_disp} but note that the normalization of this equation needs to be calibrated against simulations (and will likely change). This equation provides testable scalings for the dependence of Eq.~\ref{eq:Mach_coag} on physical parameters. We defer this to future work.

We can also use this to understand why there is a critical final overdensity $\chi_{\rm crit} \approx 300$  for recollapse and coagulation during `shattering' \citep{Gronke2020}. For an expanding cloud to achieve momentum contact with its surroundings and decelerate, it must sweep up of order its own mass: $\rho_{c} r_{\rm cl}^3 \approx \rho_{\rm h} d^{3}$, which gives $d \approx \chi^{1/3} r_{\rm cl}$. Assuming droplets are launched at a velocity $v \sim c_{\rm s,c}$, we have: 
\begin{equation}
\frac{U_{\rm tot}}{K_{\rm tot}} \sim \frac{M_{\rm h}} {M_{\rm c}} \frac{v_{\rm mix}^{2}}{c_{\rm s,c}^{2}} \left(\frac{N r_d^2}{\chi^{2/3} r_{\rm cl}^2} \right)^{2}  \propto \frac{N^{2/3}}{\chi^{4/3}}
\end{equation} 
where $A_{\rm tot} \sim N r_{\rm d}^{2} \sim r_{\rm cl}^{2} (r_{\rm cl}/r_{\rm d})$, where $N \sim (r_{\rm cl}/r_{\rm d})^{3}$. The first two factors $M_{\rm h}/M_{\rm c}$ and $v_{\rm mix}^{2}/c_{\rm s,c}^{2}$ are order unity. The number of cloudlets $N$ is difficult to model, but it is clear that as overdensity $\chi$ increases, the ratio $U_{\rm tot}/K_{\rm tot}$ decreases, and eventually coagulation is not possible. Overdense gas is lauched out to larger distances $d$ before it is slowed down by the hot gas, and by that time, the covering fraction $f_{\rm A}$ drops sufficiently that coagulation is suppressed.

Of course, more careful study and detailed comparisons to simulations are required to transform these remarks into a quantitative theory, which we defer to future work.

\subsection{The competition of coagulation versus dispersion}
\label{sec:coag_disp}

At first blush, the results of this paper might suggest that coagulation should be unimportant. The coagulation velocity $v_{\rm coag} \sim v_{\rm mix} \sim c_{\rm s,c}$ is small and diminishes rapidly with distance, $v_{\rm coag} \propto d^{-2}$. This corresponds to a small Mach number, even a relatively small distance from the cloud: 
\begin{equation} 
  {\mathcal M} \sim \frac{v_{\rm mix}}{c_{\rm s,h}} \left( \frac{r_{\rm cl}}{d}\right)^{2} \sim 10^{-2} \left( \frac{v_{\rm mix}}{c_{\rm s,c}} \right) \left( \frac{\chi}{100} \right)^{-1/2} \left(\frac{d}{3 r_{\rm cl}} \right)^{-2}
  \label{eq:Mach_coag_collection2}
\end{equation}
which would appear miniscule compared to other velocities in the system, so that coagulational inflow is a negligibly small perturbation. Yet, there are configurations, such as cloud crushing and cloud shattering, where coagulation is undeniably important. Indeed, a multi-phase mixing layer \citep{Kwak2010,Tan2020} is itself an example of coagulation -- despite the high velocity of shearing hot gas, $v_{\rm shear} \sim {\mathcal M} c_{\rm s,h} \gg v_{\rm mix} \sim c_{\rm s,c}$, cooling gas fragments in the mixing layer advect towards the cold gas layer. 

The previous section (\S\ref{sec:coag_grav}) addressed the $v_{\rm coag} \propto d^{-2}$ fall-off. This only holds for a single cloud. If surface area is enhanced (e.g, by fragmentation), so that the area covering fraction $f_{\rm A}$ is large, then the fall-off with distance is supressed. Thus, for instance, $U_{\rm tot} \sim M_{\rm h} v_{\rm mix}^{2}$ when $f_{\rm A} \sim 1$ (Eq.~\eqref{eq:U_pot}); all the hot gas is moving with velocity $v_{\rm mix}$. This is similar to Obler's paradox: if every sightline in an infinite static universe ends on the surface of a star, then the surface brightness of the night sky would be that of a stellar surface, since the reduced solid angle (which increases the number of stars which tile the sky) and inverse square dimming behave in the same way. Similarly, if $f_{\rm A} \sim 1$, then $v_{\rm coag} \sim v_{\rm mix}$, regardless of distance. Alternatively, we can use the analogy between gravity and coagulation to use Gauss's law to find how $v_{\rm coag}$ diminishes with distance. For the cometary tail of a cloud in a wind or a filamentary cold gas structure, $F_{\rm coag} \propto v_{\rm coag} \propto d^{-1}$ (as for the gravitational force of a filament). 
For a semi-infinite slab of cold gas (as in a mixing layer), $F_{\rm coag} \propto v_{\rm coag} \sim v_{\rm mix}$ is independent of distance (as for the gravitational field above a mass sheet). Slab-like geometry can arise in strongly stratified atmospheres, and filamentary geometry can arise in systems with strong B-fields. 

Still, that leaves the second question: even if $v_{\rm coag} \sim v_{\rm mix} \sim c_{\rm s,c}$, how can that compete against much larger turbulent velocities $\sigma_{\rm t} \sim \mathcal{M}_{\rm h} c_{\rm s,h}$? Indeed, it cannot in general\footnote{It is true that $\sigma \propto l^{1/3}$ in Kolmogorov turbulence, so that turbulence decreases at small scales, but $\sigma_{\rm l} < c_{\rm s,c}$ is only true for $l < (\chi^{-1/2} \mathcal{M}_{\rm h})^3 L$, where $L$ is the driving scale. For instance, for $\chi \sim 100, M_{\rm h} \sim 0.5$, then $l \lsim 0.01 L$. Such scales are at best only a few grid cells apart in simulations where coagulation is seen, and coagulating cloudlets are generally separated by larger distances.}.
However, it can in laminar bulk flows (where even if the bulk flow velocity is large, the relative velocity between cold gas fragments is small as they entrain in the hot wind), or in quiescent regions of a turbulent medium. For instance, as hot and cold gas mix, the `mass loading' of cold gas into the hot gas results in a new velocity dispersion $\sigma_{\rm mix}$, where $\langle \rho \rangle \sigma_{\rm mix}^2 \approx \rho_{\rm h} \sigma_{\rm t}^{2}$, and $\langle \rho \rangle \sim f_{\rm c} \rho_{\rm c} + (1-f_{\rm c}) \rho_{\rm h} \approx f_{\rm c} \rho_{\rm c}$. This gives $\sigma_{\rm mix} \sim \sigma_{\rm t}/(
\chi f_{\rm c})^{1/2} \sim \mathcal{M}_{\rm h} f_{\rm c}^{-1/2} c_{\rm s,c}$, so that $\sigma_{\rm mix} < c_{\rm s,c}$ if $f_{\rm c} >  \mathcal{M}_{\rm h}^2$. Note that here $\sigma_{\rm mix}$ is the velocity dispersion of the multi-phase (hot and cold) gas mixture; it is {\it not} the velocity dispersion of mixed gas at some intermediate temperature. All situations where coagulation is observed to be important (e.g., coagulation onto the cometary tail of a cloud; shattering; turbulent mixing layers) are those where cold gas mass loading $f_{\rm c}$ is fairly large and the gas turbulent velocity {\it does} in fact obey $\sigma_{\rm mix} \lsim v_{\rm mix}$. 
The same is not true if $\langle \rho \rangle \ll \rho_{\rm c}$, and thus $\sigma_{\rm mix} > v_{\rm mix}$. For instance, in the driven turbulence multi-phase setup of \citep{Gronke2022} (where $\sigma_{\rm mix} > v_{\rm mix}$; see fig 19 in that paper), coagulation indeed does not outcompete fragmentation by turbulence. Instead of coalescing into a large central cloud, a scale free power law mass distribution of clouds forms. 

To summarize: coagulation is efficient, despite the small amplitude $v_{\rm coag} \sim v_{\rm mix} \sim c_{\rm s,c}$ and rapid fall-off $v_{\rm coag} \propto d^{-2}$, in regions where \textit{(i)} the extrinsic dispersion velocity is low, e.g., if the cold gas fraction $f_{\rm c}$ is high, and $\langle \rho \rangle \approx f_{\rm c} \rho_{\rm c}$ (since this mass loading reduces the turbulent velocity to $\sigma_{\rm mix} < c_{\rm s,c}$ if $f_{\rm c} > \mathcal{M}_{\rm h}^2$) and \textit{(ii)} the geometrical dimming can be overcome, for instance, through the geometry of the cold medium or if the cold gas covering fraction $f_{\rm A}$ is high (since the fall-off with distance in $v_{\rm coag}$ goes away as $f_{\rm A} \rightarrow 1$). \\

Next, we discuss some of these cases where coagulation is important in more detail:
\begin{itemize}
    \item{{\it Mixing layers, clouds and streams.} Plane parallel Kelvin Helmholtz mixing layers have $f_A \sim 1$, and thus $v_{\rm coag} \sim v_{\rm mix}$ does not decline with distance. Also, regions where cold gas mass loading is substantial have turbulent velocities $u^{\prime} \sim v_{\rm shear}/\sqrt{\chi} \sim c_{\rm c,s}$, thus comparable to $v_{\rm mix}$, as one might expect from the above arguments. Clouds in a hot wind develop an extended cometary tail with a cylindrical structure \citep[e.g.][]{Gronke2019}. Thus, entrained clouds, or cold gas streams \citep[e.g.,][]{mandelkerInstabilitySupersonicCold2020,Bustard2022} correspond to our 2D (Fig \ref{fig:coag2d_veldrop_multiplot}), rather than our 3D (Fig \ref{fig:coag3d_multiplot}) simulations, with $v_{\rm in} \propto d^{-1}$ instead of $v_{\rm in} \propto d^{-2}$. Similar to Fig \ref{fig:coag2d_veldrop_multiplot}, the droplet returns on a timescale $\sim \tilde{\alpha} t_{\rm sc,floor}$ (where $\tilde{\alpha} \sim 5-10$), during which time it travels a distance $\sim v_{\rm w} \tilde{\alpha} t_{\rm sc,floor} \sim \tilde{\alpha} \chi^{1/2} \mathcal{M} r_{\rm cl}$.}

\item {\textit{Expulsion from a central origin.} 
Droplets dispersed from a central origin can eventually coagulate back together. In \citet{Gronke2020}, we argued that the competition between dispersion and coagulation sets the threshold of `shattering' which we found to be $\chi_{\rm final}\gtrsim 300 (r_\cl / 10^4 \lshatter)^{1/6}$  (for $r_\cl\gg \lshatter$ and $\delta P / P \gtrsim 1.5$). In our simulations, clouds straddling this boundary had vastly different outcomes.
 In principle, since drag forces cause kinetic energy to decay, coagulation could potentially once again dominate at late times\footnote{Similarly, pulsations could potentially damp in an otherwise static medium. However, we showed that they continue on timescales much longer than $t_{\rm sc}$, and in practice turbulence will always perturb the cloud.}, though in practice turbulence will further separate the fragments and shape the mass distribution.  
  While the exact mechanism of fragmentation and dispersion in the `shattering' scenario needs revisiting, in broad terms the role of coagulation here is clear.} 

\item {\textit{Coagulation in extrinsic turbulence.} In \citet{Gronke2022}, we studied turbulent, multiphase dynamics in more detail and found that the droplets follow a power law mass distribution $\dd n/\dd m \propto m^{-2}$ (which is also found in larger scale simulations of the intracluster medium; cf. \citealp{LiMODELINGACTIVE2014a}). This simulations were all run at Mach numbers above the critical Mach number (Eq.~\eqref{eq:Mach_coag_numerical}) where we might expect coagulation to play a role. For low Mach numbers, we expect deviations from this power law, which we will analyze in future work.}

\end{itemize}

Note that in the above scenarios mixing and subsequent cooling is often not necessarily provided by the `cooling induced pulsations' discussed in \S~\ref{sec:convergence_mdot}; all that is necessary is that mixing takes place.

\subsection{Caveats}
\label{sec:disc_caveats}

Our study does not address a range of topics which we hope to revisit in future work.
\begin{itemize}
\item Magnetic fields. Most plasmas are magnetized, which affects the mixing and thus the mass transfer process \citep{Ryu1995,Ji2018}. Furthermore, $B$-fields imply a non-thermal pressure support which can become large in the cold medium even with initially large plasma $\beta$ due to magnetic compression \citep{Sharma2010,Gronke2019}.
\item Cosmic rays. Similar to magnetic fields, cosmic rays are present in astrophysical plasmas and provide non-thermal pressure support which changes the cooling rates of the gas \citep[e.g.,][]{2016MNRAS.456..582S,Butsky2020}.
\item Simplified setup. The goal of this study was to develop a simple model for coagulation. Thus, we focused on very simplified initial conditions. In future work, we want to apply this model in more realistic scenarios such as multiphase galactic winds.
\item Resolution and dynamic range. As all numerical studies, we suffer from finite resolution and limited dynamic range. We tried to support our findings with resolution studies throughout as well as analytic models matching our numerical findings. Note that throughout this work we do not aspire to achieve full convergence, i.e., to have converged cold gas morphology. We merely aim for convergence in cold gas mass.
\end{itemize}

We do not consider the neglect of thermal conduction to be a major caveat of this work. We have previously shown that when heat diffusion is dominated by turbulent mixing (as is true here), thermal conduction has little impact on mass growth rates and coagulation velocities \citep[see section 4.6, 5.5 and 5.6 in][for a detailed discussion]{Tan2020}.

\subsection{Comparison to the literature}
\label{sec:comparison_literature}

The oscillations for cooling clouds were previously discussed in the literature, in particular in \citet{Waters2019b} and \citet{Das2020} in 1D and in \citet{Gronke2020} in 3D simulations. Notably, \citet{Waters2019b} and \citet{Das2020} carry out in-depth analyses of linear, non-isobaric thermal instability and 
found pulsations for `large clouds'.
\citet{Waters2019b} show that the pulsations decay on a long ($\gtrsim 10 t_{\rm sc}$) timescale, and that the cloud settle eventually ($\gtrsim 50 t_{\rm sc}$) back to the equilibrium state. However, note that gas mixing -- which can fuel cooling and further pulsations -- is not captured in 1D.

The further mass growth associated with these pulsations was not studied in these simplified setups. In \citet{Gronke2018}, similar pulsations were seen in the entrained state of a `windtunnel' simulations. There, the pulsations were, thus, not seeded by the initial cooling but by the shock hitting the cloud. The mass growth rates of the cold gas agree well including the characteristic $\propto t_{\rm cool}^{-1/4}$ scaling. The rates as well as the scalings have been confirmed in turbulent mixing layer simulations 
\citep{Fielding2020,Tan2020}\footnote{Note that, more recently, these scalings were extended into the high-$\mathcal{M}$ regime (\citealp{Yang2022}; see also \citealp{Bustard2022}).}, although there gas mixing is driven by shear, rather than pulsations from overstable sound waves.

Similarly, coagulation was observed in previous studies. \citet{ZelDovich1969} computed the cooling rate at the (laminar) boundary of a two-phase medium, and noted that this leads to coagulation of the cold medium. However, they also point out that this velocity of the front is minuscule. Building upon this work, \citet{Elphick1991} constructed a 1D framework to study the coagulation of an ensemble of cold gas fronts -- which they extend to include bulk fluid motions in \citet{Elphick1992}. This work was later extended to more dimensions \citep{Shaviv1994}.

More recently, \citet{Koyama2004} and \citet{Waters2019} also discuss cooling induced coagulation in one- and two-dimensions, respectively. They point out that not only do the fronts move due to growth of the cold gas, but that motion is induced by the cooling (see, e.g., Fig. 2 in \citealp{Koyama2004}). In particular, \citet{Waters2019} analyze the coagulation behavior and note the coalescence timescales.

Our work differs from these previous studies in several aspects. Firstly, we carry out two- and three-dimensional simulations and build an analytic model reproducing our hydrodynamical results. More importantly, however, we focus on the production of intermediate temperature gas by turbulent mixing--which cannot be captured in 1D--as opposed to laminar heat transport due to thermal conduction alone.  For this reason, our coagulation velocities are much greater (in the same way as the turbulent diffusion dominates over laminar heat transport; cf. \citealp{Tan2020}). For instance, for their fiducial 2D run in which they placed a $r \sim 8\lshatter$ and a $r \sim 23 \lshatter$ clouds at a distance of $d\sim 33 \lshatter$, \citet{Waters2019} found a coagulation velocity of order $v\sim d/t_{\rm coag}\sim 0.02 c_{\rm s}$ (cf. their table 4) 
which is much less than the $v_{\rm coag}\sim v_{\rm mix} r/d \sim c_{\rm s}$ we find\footnote{Note that the small cloud sizes they employed would lead to reduced pulsations and thus an actual slower coagulation velocity; see Fig.~\ref{fig:mdot_vs_X}.}.

Coagulation can be observed in many larger scale simulations. As already mentioned, many `cloud crushing' simulations with radiative cooling display signs of cold gas coagulation \citep[e.g.,][]{Schneider2016,Gronke2019,Abruzzo2021}. Also in simulations studying thermal instabilities, coagulation has been observed \citep{Sharma2010,Butsky2020}.
In even larger scale simulations, e.g., of the multiphase dynamics in the CGM \citep[e.g.][]{Hafen2019,Hummels2018}, the ICM \citep{LiMODELINGACTIVE2014}, or in ram pressure stripped tails of galaxies \citep{Tonnesen2010,Farber2022} coagulation should take place and play a role -- however, it is unclear whether current resolutions are sufficient to capture this effect.

\section{Conclusion}
\label{sec:conclusion}
We investigated cooling driven coagulation process of cold gas in a multi-phase 
medium, a phenomenon which has been seen in diverse simulations. For instance, it is observed in the `focusing' of cold gas droplets onto the cometary tail of a cold cloud in a hot wind. To gain understanding, we first investigated cooling induced coagulation in a static medium.
Our findings can be summarized as follows:
\begin{enumerate}
\item Perturbed cold gas blobs of size $> \lshatter \equiv \mathrm{min}(c_{\rm s}t_{\rm cool})$ develop pressure fluctuations which lead to continuous pulsations and mass growth of the cold gas.
\item This process leads to a flow of hot gas with velocity $v_{\rm coag} \sim v_{\rm mix} (r_{\rm cl}/d)^2$ in 3D, where $v_{\rm mix}$ (given by Eq.~\eqref{eq:vmix}) is of order the cold gas sound speed. Cold droplets can get rapidly entrained in this hot gas flow (due to their own growth and the associated momentum transfer), eventually merging with other cold gas structures. 
\end{enumerate}
We furthermore developed an analytic model describing the mass growth and coagulation process which fits our numerical results reasonably well. Although $v_{\rm coag} \sim v_{\rm mix} (r_{\rm cl}/d)^2$ may appear small, note that: (i) turbulent gas velocities can be small, e.g., if the cold gas mass fraction is high. Also, in bulk flows (as in a wind), the relative velocities between entrained gas fragments becomes small. (ii) The geometric $v_{\rm coag} \propto d^{-2}$ dimming can be much weaker in different geometries (e.g., $v_{\rm coag} \propto d^{-1}$ for a cometary tail), or if the cold gas covering fraction $f_A$ is high (where $v_{\rm coag} \approx v_{\rm mix} \approx$const).

Our finding supports the idea that cooling driven coagulation of adjacent cold gas is possible and we establish a criterion defining the regimes where coagulation or dispersion in transonic turbulence dominates.
Due to the similar $F \propto d^{-2}$ force, we draw analogies to gravity. The monopole term for coagulation is surface area, rather than mass. Thus, fragmentation, which increases area at fixed mass, also increases  coagulation. 

We have neglected magnetic fields, cosmic rays, and the inclusion of a more realistic (turbulent and stratified) background. We plan to address these issues in future work.

\section*{Acknowledgments}
We thank both the referee and Nir Mandelker for detailed comments which improved the quality of this work. This research made use of \texttt{yt} \citep{2011ApJS..192....9T}, \texttt{matplotlib} \citep{Hunter:2007}, \texttt{numpy} \citep{van2011numpy}, and \texttt{scipy} \citep{scipy}.
We acknowledge support from NASA grant NNX17AK58G, HST grant HST-AR-15039.003-A, and XSEDE grant TG-AST180036 the Texas Advanced Computing Center (TACC) of the University of Texas at Austin.
MG thanks the Max Planck Society for support through the Max Planck Research Group.
 This research was supported in part by the National Science Foundation under Grant No. NSF PHY-1748958.

\section*{Data Availability}
Data related to this work will be shared on reasonable request to the corresponding author.

 \bibliographystyle{mnras}
 
 \bibliography{references_linked,references_downloaded}

\appendix

\section{Convergence of mass growth}
\label{sec:convergence}
\subsection{Importance of pulsations for convergence in mass growth}
\label{sec:osci_conv}

\begin{figure}
  \centering
  \includegraphics[width=\linewidth]{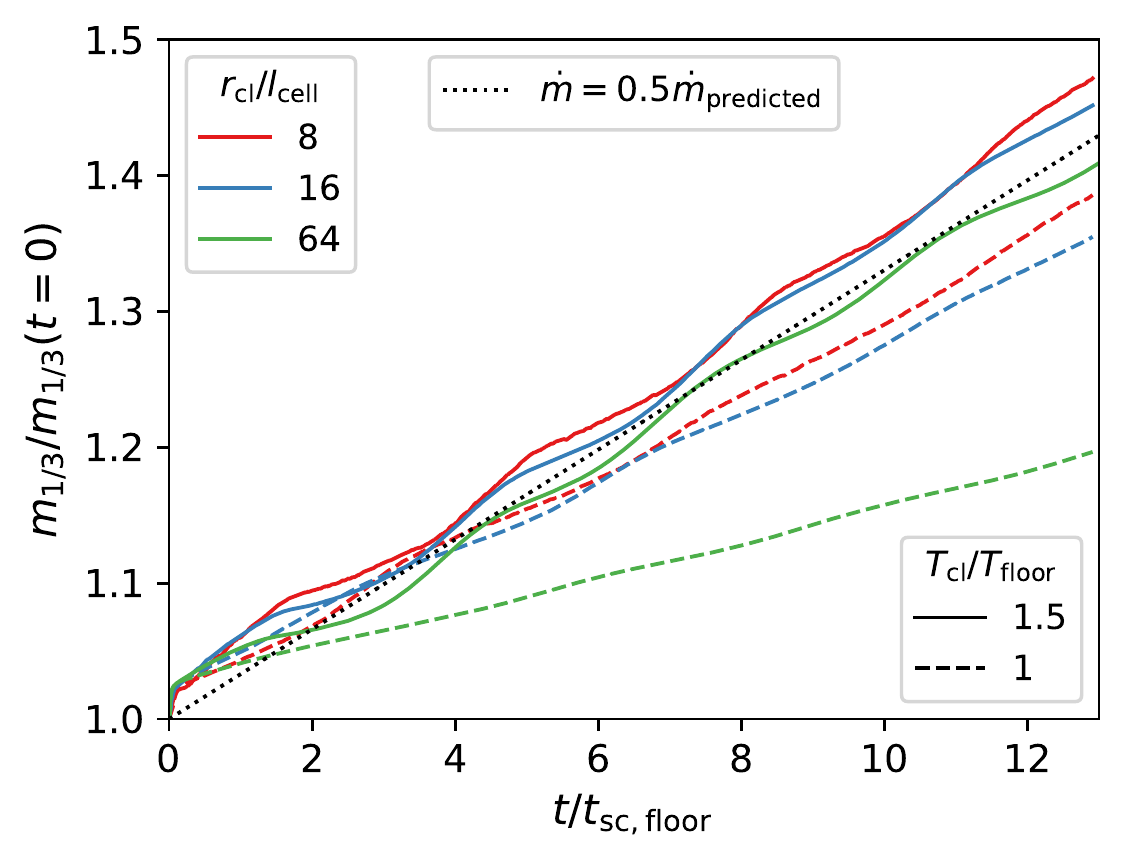}
  \caption{Mass evolution of a perturbed and non-perturbed (solid and dashed lines, respectively) cold gas blob using various resolutions (marked with different colors). The dotted line shows the approximate mass growth using Eq.~\eqref{eq:mdot}.
    }
  \label{fig:convergence_Tcl1}
\end{figure}

\begin{figure*}
  \centering
  \includegraphics[width=.49\linewidth]{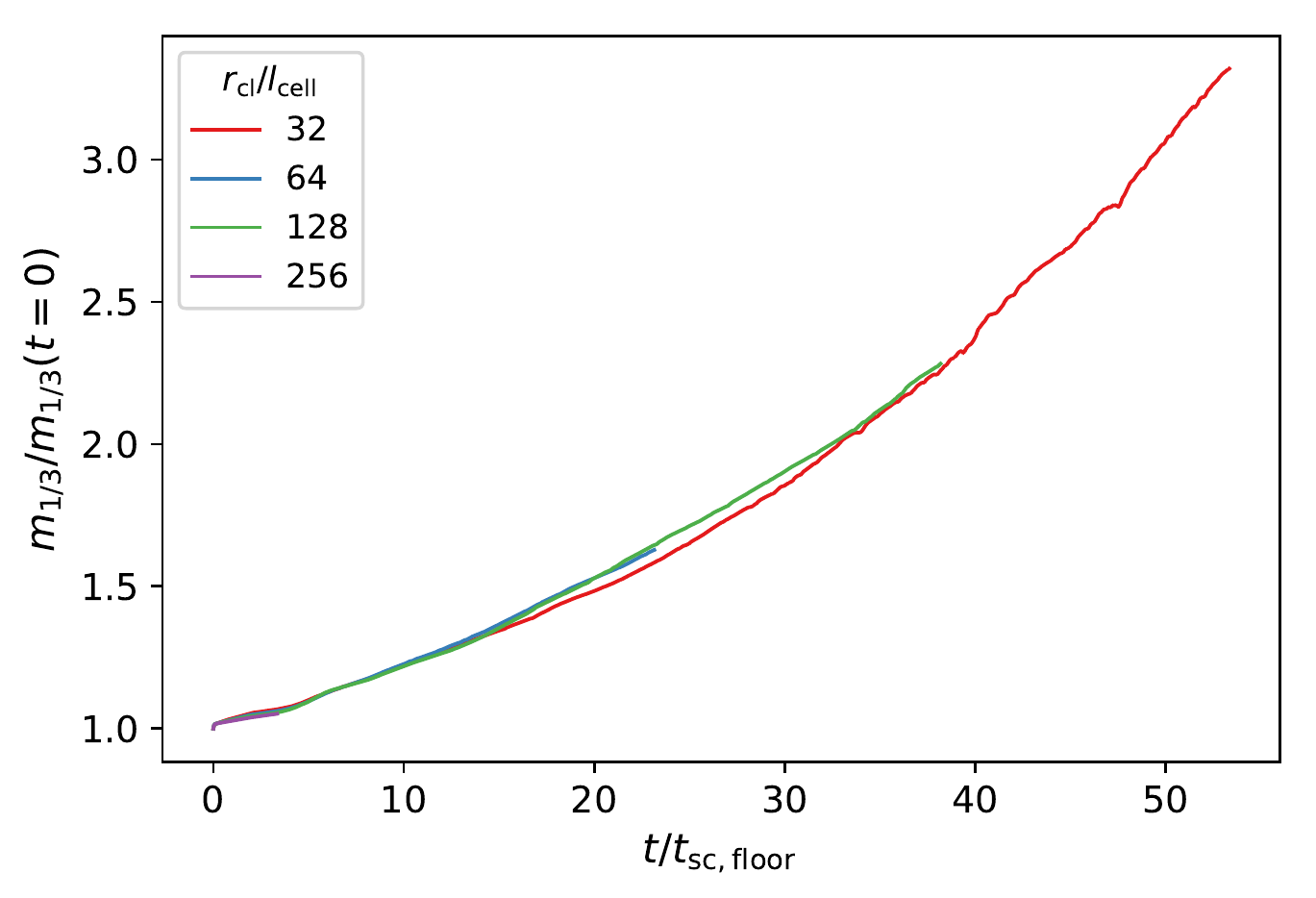}
  \includegraphics[width=.49\linewidth]{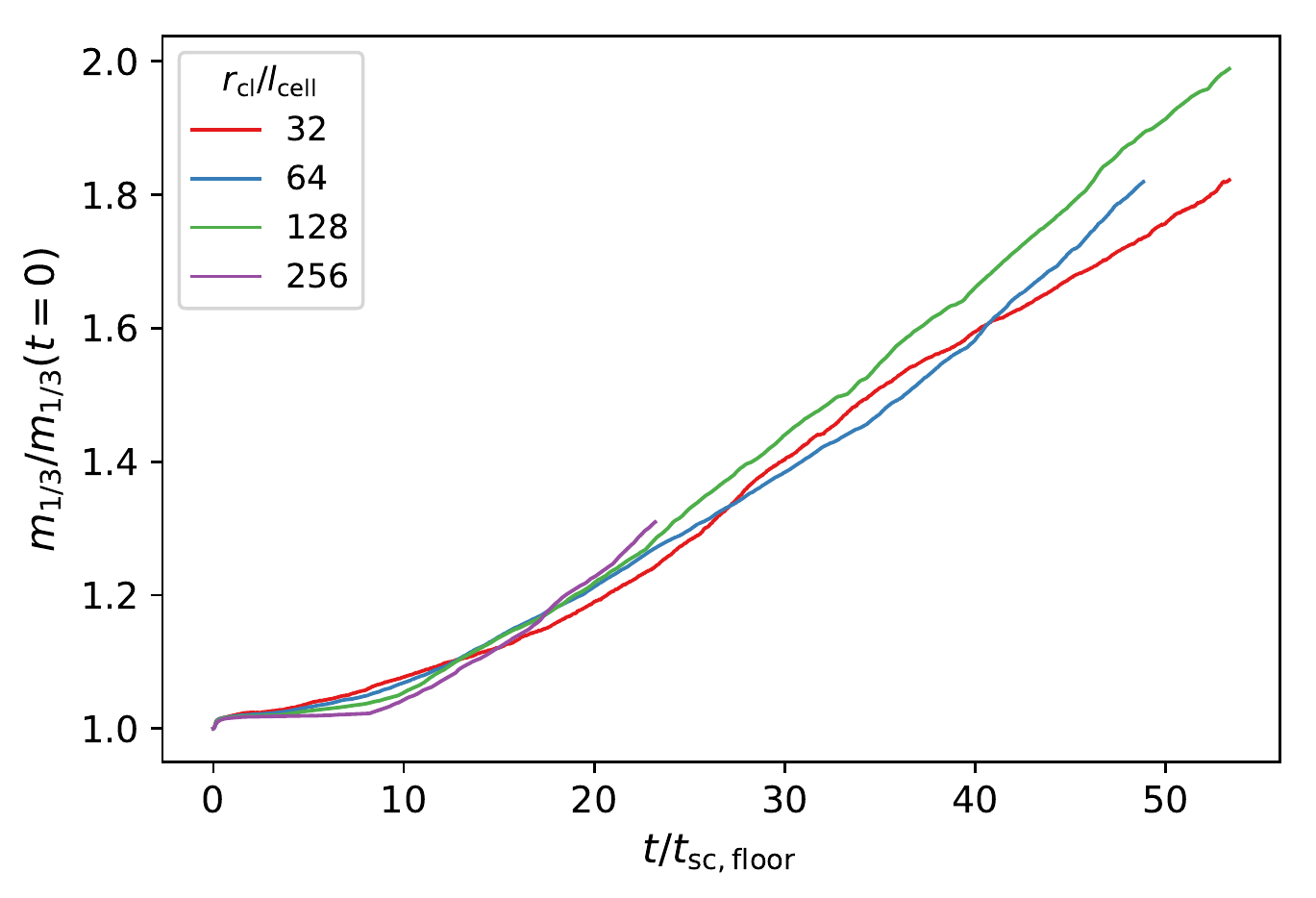}
  \caption{2D convergence study of pulsating blobs. The \textit{left panel} shows our fiducial setup ($\chi\sim 100$, $T_\cl / T_{\rm floor}=1.5$, $r_\cl / \lshatter \sim 2500$). In the \textit{right panel} the clumps are a factor $40$ smaller, i.e., $r_\cl / \lshatter \sim 60$ allowing us to resolve $\lshatter$.}
  \label{fig:conv2d_pulsations}
\end{figure*}

\begin{figure}
  \centering
  \includegraphics[width=\linewidth]{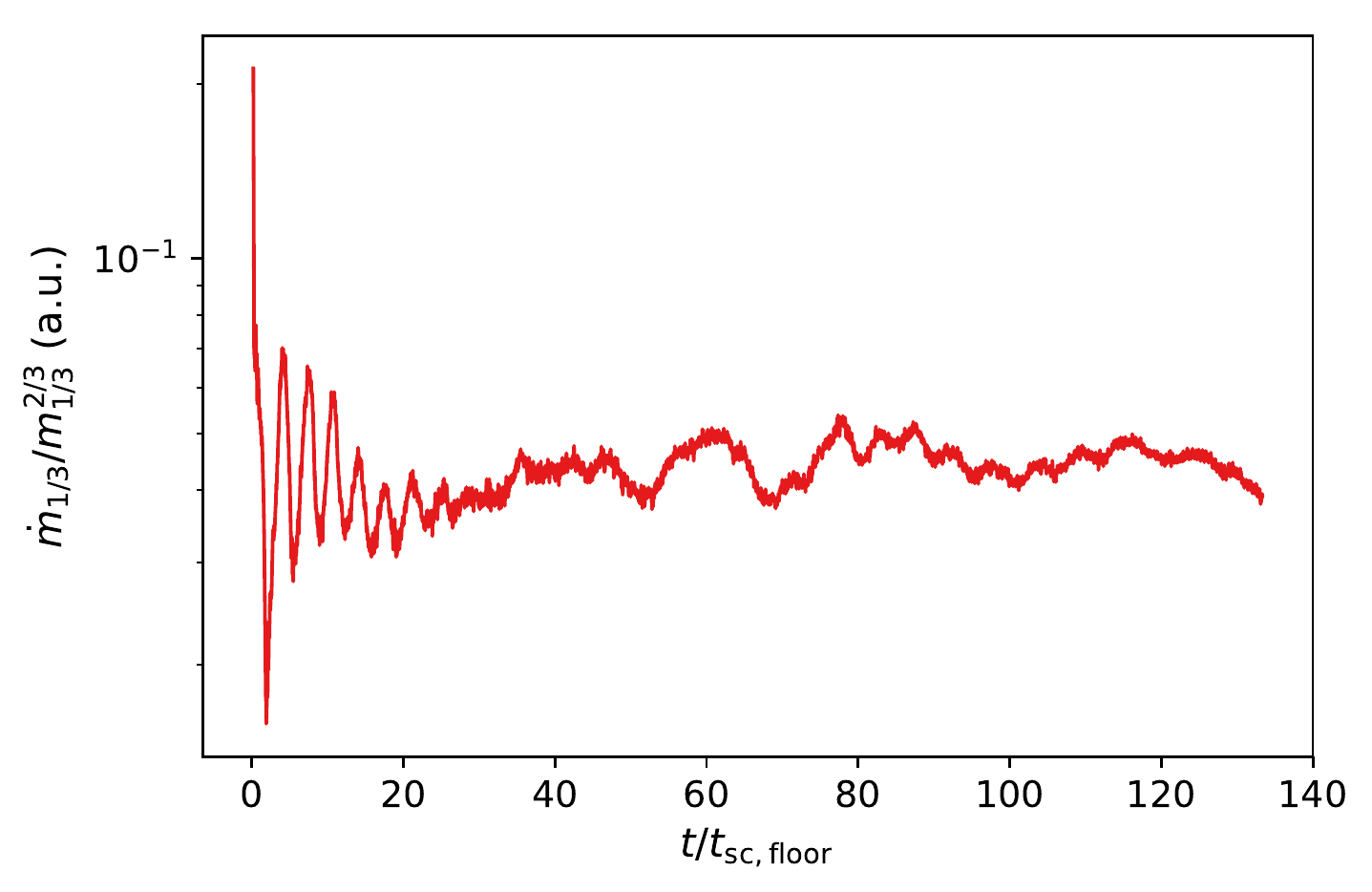}
  \caption{Mass growth for our fiducial $\chi\sim 100$, $r_\cl / \lshatter \sim 2500$, $r_\cl / l_{\rm cell}\sim 16$, $T_\cl / T_{\rm floor}\sim 1.5$ run for an extended period of time. Clearly the pulsations and the mass growth does not cease.}
  \label{fig:mdot_longrun}
\end{figure}

Figure~\ref{fig:convergence_Tcl1} shows the cold gas mass evolution for our 3D `static' setup ($\chi\sim 100$, $r_\cl / \lshatter \sim 2500$) with different resolutions ranging from $8$ to $64$ cells per cloud radius. The solid lines indicate the evolution for a perturbed cloud which are \textit{(i)} fairly well converged, and \textit{(ii)} follow the expected behavior from Eq.~\eqref{eq:mdot} while the dashed lines show unperturbed clouds in which the mass growth is driven by numerical diffusion, and is unconverged. Note that we also show different resolution runs in Fig.~\ref{fig:mdot_vs_rcl_overview} and Fig.~\ref{fig:mdot_vs_X} for different overdensities and cloud sizes showing that the mass growth rates (for pulsating clouds) are approximately resolution independent.

Figure~\ref{fig:conv2d_pulsations} shows the same convergence test but in 2D and (in the right panel) with a smaller cloud of size $r_\cl / \lshatter\sim 60$. Thus, in the highest resolution runs there, $\lshatter$ is resolved. Still, the growth rates are similar to that in runs where $\lshatter$ is not resolved.

Figure~\ref{fig:mdot_longrun} shows the same convergence test as in Fig.~\ref{fig:convergence_Tcl1} but for an extended period of time ($>100\,t_{\rm sc,floor}$). We can see that the mass transfer rate does not decay but instead continues to grow as $\dot m\propto A_{\rm cl}\propto m^{2/3}$ (cf. Eq.~\eqref{eq:mdot}) as expected from monolithic growth. While the initial pulsation pattern is imprinted for $\sim 30\,t_{\rm sc}$, more unstructured pulsations dominate later on leading to continuous mass growth.

\subsection{Change of box size}
\label{sec:Lbox_conv}

\begin{figure}
  \centering
  \includegraphics[width=\linewidth]{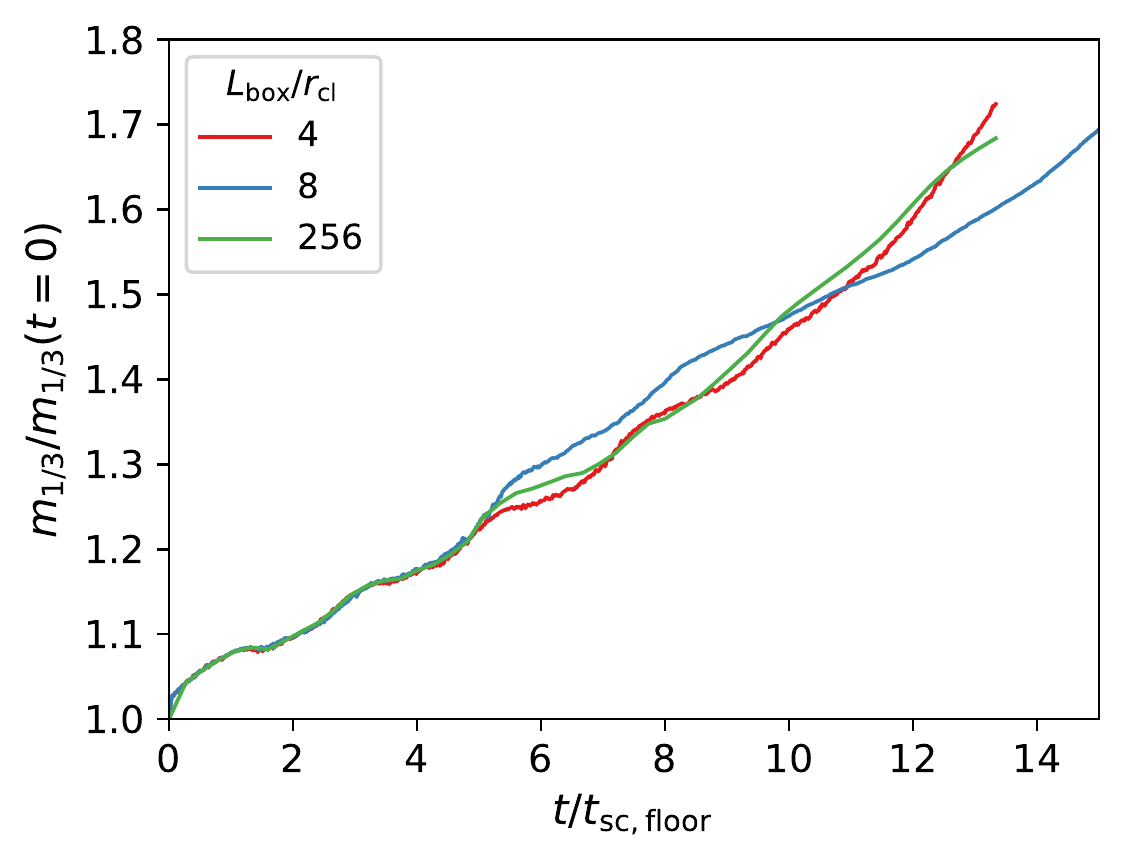}
  \caption{Mass evolution for clouds with $T/T_{\rm floor}\sim 2$, $r_{\rm cl} / l_{\rm cell}=16$ for different box sizes. Note that for the largest box, we used a static refined mesh for the inner region.}
  \label{fig:massevo_boxsize}
\end{figure}

We checked whether the pulsations are caused by reflecting waves off the simulation boundary by increasing the boxsize.
Figure~\ref{fig:massevo_boxsize} shows the mass evolution for three box sizes. For the largest box size we used a statically refined mesh for the inner region with side length $\sim 8 r_{\cl}$.
Since we have 
$t_{\rm sc,box}\sim L_{\rm box}/ c_{\rm s,h} \sim L_{\rm box}/(\chi^{1/2} c_{\rm s,c}) \sim (L_{\rm box}/r_{\rm cl})(t_{\rm sc,floor}/\chi^{1/2}) \sim 25 t_{\rm sc,floor}$ for our largest box where $L_{\rm box}/r_{\rm cl}$, in our largest box, reflecting waves cannot perturb the cloud (over a run time of $\sim 14 t_{\rm sc,floor}$), but the mass growth is consistent with smaller boxes.

\subsection{Convergence test for coagulation}
\label{sec:conv_coag}

\begin{figure}
  \centering
  \includegraphics[width=\linewidth]{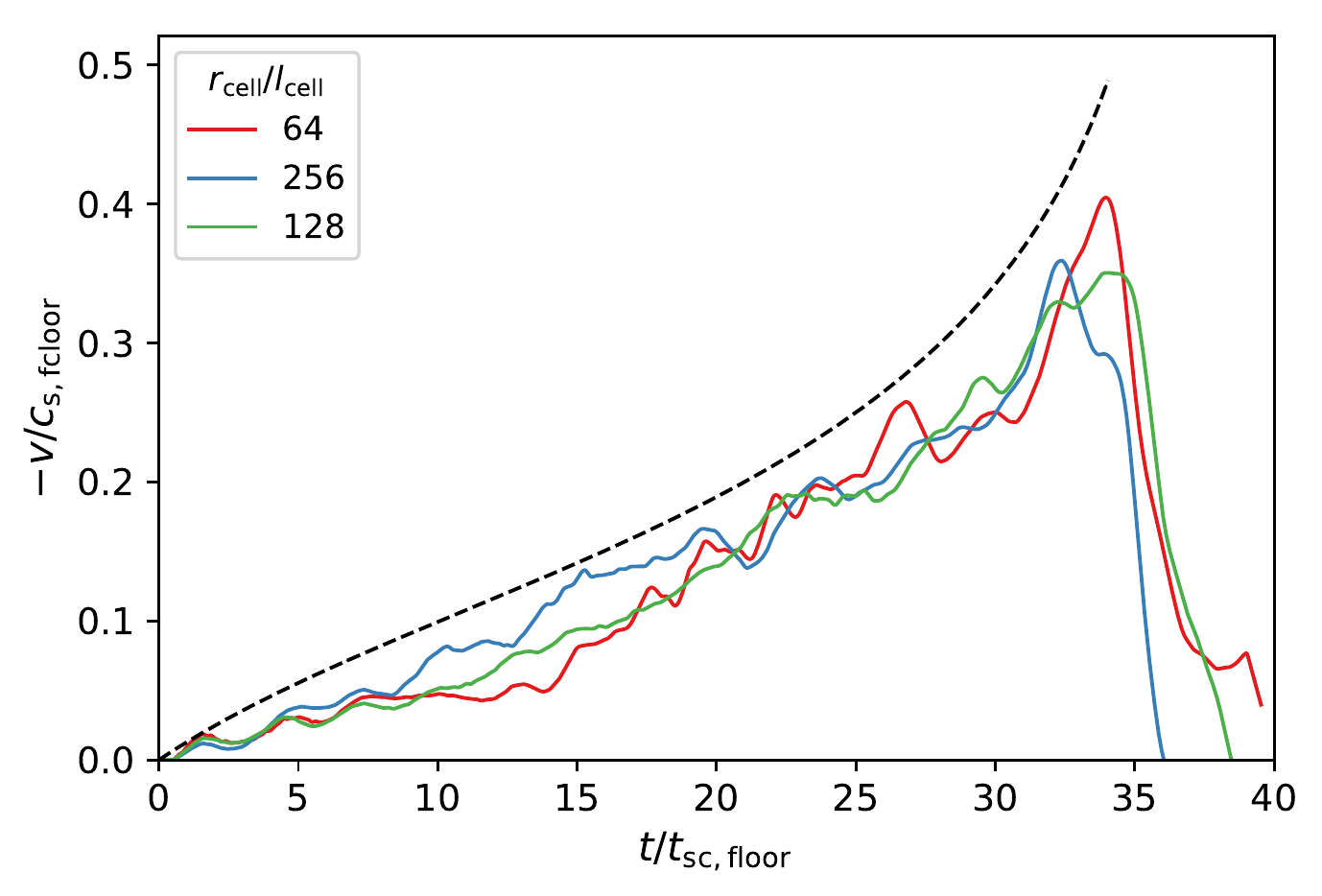}
  \caption{Convergence test for the coagulation process. The setup is the same as in Fig.~\ref{fig:coag2d_multiplot}/\S~\ref{ssec:2dcoag} with $d_0/r_\cl =8$, $T_\cl/T_{\mathrm{floor}}=2$.}
  \label{fig:coag_convergence}
\end{figure}
Figure~\ref{fig:coag_convergence} shows our coagulation setup discussed in Fig.~\ref{fig:coag2d_multiplot}/\S~\ref{ssec:2dcoag} with different resolutions and shows the coagulation process is fairly converged.

\bsp	
\label{lastpage}

\end{document}